\documentclass[showpacs,apl,twocolumn,floatfix]{revtex4-1}

\usepackage{graphicx,float,amsmath}
\usepackage{ mathrsfs }
\usepackage{natbib}

\begin{document}

\date{\today}

\title{Effect of the Pauli principle on photoelectron spin transport in $p^+$ GaAs}

\author{F.~Cadiz$^1$}
\author{D.~Paget$^1$}
\author{A. C. H. ~Rowe$^1$}
\author{T.~Amand$^2$}
\author{P. ~Barate$^2$}
\author{S.~Arscott$^3$}

\affiliation{%
$^1$Physique de la Mati\`ere Condens\'ee, Ecole Polytechnique, CNRS, 91128 Palaiseau, France}

\affiliation{%
$^2$Universit\'e de Toulouse, INSA-CNRS-UPS, 31077 Toulouse Cedex, France}

\affiliation{%
$^3$Institut d'Electronique, de Micro\'electronique et de Nanotechnologie (IEMN), Universit\'e de Lille, CNRS, Avenue Poincar\'e, Cit\'e Scientifique, 59652 Villeneuve d'Ascq, France}

\begin{abstract}
In $p^+$ GaAs thin films, the effect of photoelectron degeneracy on spin transport is investigated theoretically and experimentally by imaging the spin polarization profile as a function of distance from a tightly-focussed light excitation spot. Under degeneracy of the electron gas (high concentration, low temperature), a dip at the center of the polarization profile appears with  a polarization maximum at a distance of about $2 \; \mu m$ from the center. This counterintuitive result reveals that photoelectron diffusion depends on spin, as a direct consequence of the Pauli principle. This causes a concentration dependence of the spin stiffness while the spin dependence of the mobility is found to be weak in doped material. The various effects which can modify spin transport in a degenerate electron gas under local laser excitation are  considered.  A comparison of the data with a numerical solution of the coupled diffusion equations reveals that ambipolar coupling with holes increases the steady-state photo-electron density at the excitation spot and therefore the amplitude of the degeneracy-induced polarization dip. Thermoelectric currrents  are predicted to depend on spin under degeneracy (spin Soret currents), but these currents are negligible except at very high excitation power where they play a relatively small role. Coulomb spin drag and bandgap renormalization are negligible due to electrostatic screening by the hole gas. 
 
\end{abstract}
\pacs{}
\maketitle

\section{Introduction}
\label{intro}
Recently, a number of novel phenomena occurring during spin-dependent transport in semiconductors have been reported, including the spin Hall effect \cite{wunderlich2010}, the inverse spin Hall effect \cite{jungwirth2012}, the spin Coulomb drag effect \cite{weber2005}, and the spin helix  \cite{weber2009}. These phenomena are of interest in-and-of themselves and also because they may affect the operation of a large number of proposed semiconductor spintronic devices \cite{datta1990, wunderlich2010, zutic2007, gerhardt2011}. Experimental investigations including the use of novel techniques such as spin gratings \cite{cameron1996} or spin noise \cite{zapasskii2013} reveal that these phenomena arise from one of two possible coupling mechanisms, either spin-charge or spin-spin couplings. In the former, the spin-orbit interaction plays a central role and gives rise to the extrinsic spin Hall effect and the spin helix, as well as providing the basis for the electrical manipulation of spin \cite{wang2013,wang2013b,balocchi2011}. Spin-spin coupling, on the other hand, results in spin Coulomb drag \cite{weber2005} and a spin-dependent density of states via bandgap renormalization \cite{takahashi2008}. Recently a new spin-charge coupling phenomenon resulting from Pauli blockade in a degenerate electron gas was revealed \cite{cadiz_prl2013}, resulting in  a spin-dependence of the diffusion constant as large as $50\%$. Pauli blockade had been implicitely included in some theoretical treatments of spin polarized electron transport \cite{takahashi2008,vignale2002,flatte2006}, but had not yet been explicitely detailed nor experimentally demonstrated. It is of importance since it will naturally modify all other coupling phenomena in the degenerate limit.\ 

Here, we present a theoretical and experimental investigation of the effect of degeneracy on  spin transport  in $p^+$ GaAs using a polarized microluminescence method in which the spin polarization is measured as a function of distance from a local, diffraction-limited excitation spot \cite{favorskiy2010,paget2012,cadiz2013}. This study reveals that the dominant effect of degeneracy is the spin dependence of diffusion. Ambipolar coupling to the photo-created hole distribution is of central importance for the observation of the effects  since it acts to locally increase the electron density near the excitation spot and therefore  to increase the degree of degeneracy. A detailed theoretical analysis allows us to predict two other spin-dependent transport effects induced by degeneracy. These effects are i) spin-dependent thermoelectric currents (spin-Soret effect) \cite{brechet2010} caused by the radial temperature gradients. ii) Spin dependence of the mobility, which is strongly decreased by hole screening of the electron collisions with charged impurities in the $p^+$  material considered here. These two effects are shown, using an extensive sample characterization, to be negligible here and their demonstration requires specific experimental configurations and doping levels. Coulomb spin drag and spin-dependent bandgap renormalization effects are also negligible because of electrostatic screening by the majority holes. Finally, it is shown that the usual spin grating technique \cite{cameron1996} is not adapted to the observation of Pauli blockade coupling phenomena since only spin concentration gradients are created whereas both spin and charge concentration gradients are necessary.

The structure of the paper is as follows. The experimental section (Sec. II) contains a description of the method, a presentation of the results and a semi-quantitative interpretation. The theory presented in Sec. III considers spin transport in a semiconductor under local light excitation, for which the charge and spin densities, as well as the temperature, vary as a function of space. Sec. IV describes the relative efficiencies of the various possible mechanisms for spin transport, while the  quantitative interpretation of the results is presented in Sec. V. 

\section{Experimental}
\label{experiment}

\subsection{Method}
\begin{figure}[tbp]
\includegraphics[width=8cm, trim=1 1 1 250, clip=true] {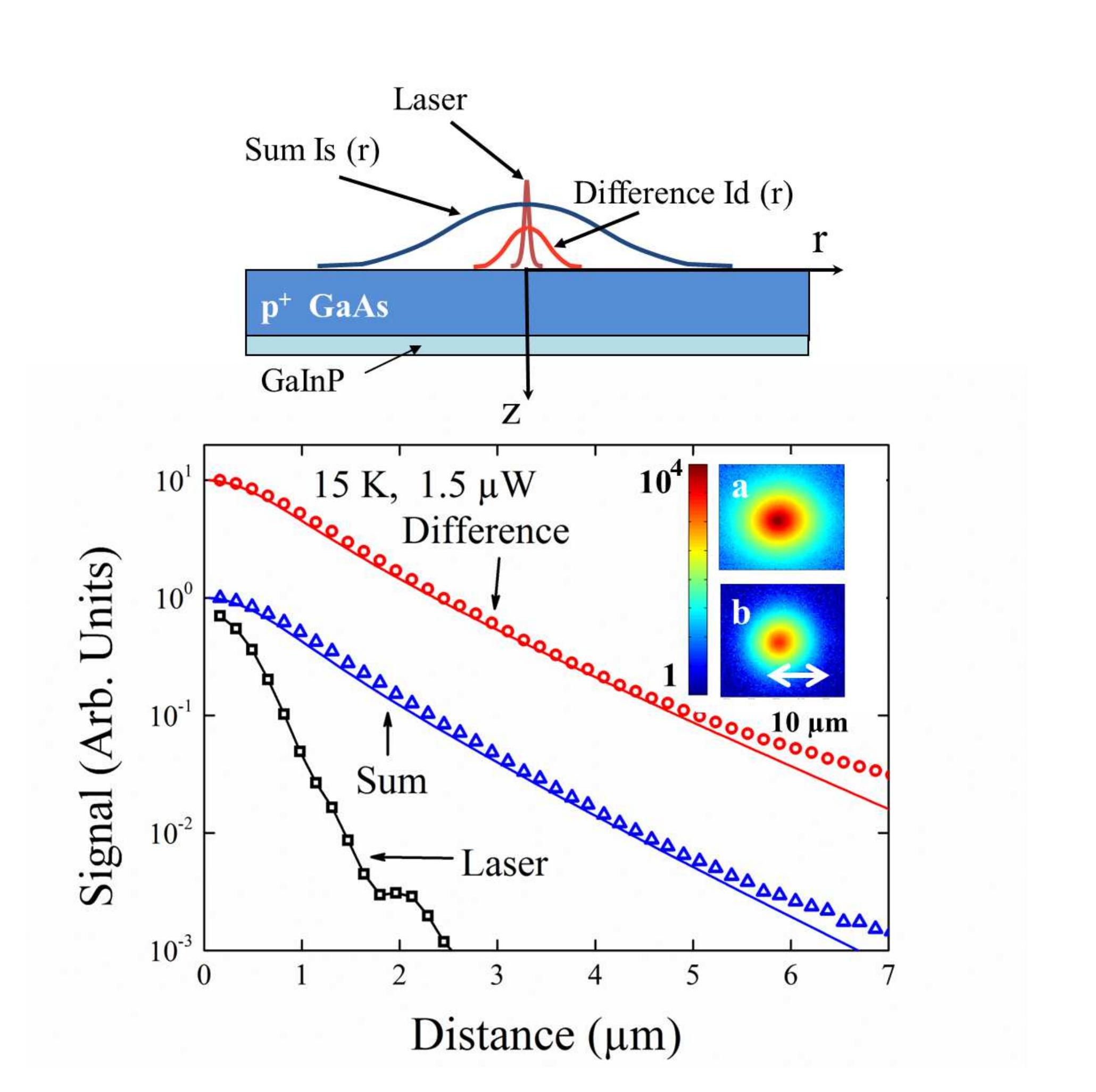}
\caption{Top panel: Principle of the experiment on $p^+$ GaAs passivated at the bottom  surface by a thin GaInP layer, with a naturally oxidized front surface. The circularly-polarized laser is tightly-focused (Gaussian radius 0.45 $\mu$m) on the GaAs film and the spatial distribution of the luminescence  sum signal [Eq.  (\ref{imagesum}), image a] and difference signal [Eq.  (\ref{imagediff}, image b] are imaged using a modified commercial microscope. The ratio of the two profiles, not shown here, gives the spin polarization profile [Eq.  (\ref{imagepol})].  The figure shows angular-integrated cross-sections of the  laser beam profile and of the above images, shifted vertically for clarity, fitted with numerical solutions of the diffusion equations [Eq. (\ref{J2}) and Eq. (\ref{J3})], that yield   estimates of the effective diffusion lengths $L_e^{eff}$ and $L_s^{eff}$. }
\label{figexp}
\end{figure}

For the experimental investigation of charge and spin diffusion, we have used $p^+$ GaAs films of thickness $d = 3\;\mu$m, grown on a GaAs semi-insulating substrate with, as shown in the top  panel of Fig. \ref{figexp}, a thin GaInP back layer to confine the photoelectrons and to ensure a negligible recombination velocity $S' = 0$ at the bottom GaAs surface. The  top surface is naturally oxidized. 

The principle of the experimental technique is shown in Fig. \ref{figexp} and has been presented in more detail elsewhere \cite{favorskiy2010}. Circularly-polarized light excitation at $1.59$ eV is focused to a Gaussian spot of  half width $\omega=0.6$ $\mu$m.  The photoluminescence (PL) of the sample only comes from the layer since emission of the semi-insulating substrate is negligible. One measures the PL intensity profile for which the  cross section as a function of radial distance $r$ from the excitation spot is related to the electronic concentration $n(r, z)$ by 
\begin{equation} \label{imagesum} I_s(r) = A\int_0^d n(r,z)\exp\left[-\alpha_lz\right] \mathrm{d}z 
\end{equation}  
where $A$ is a proportionality constant and $\alpha_l \approx (3 \mu$m$)^{-1}$ is the absorption coefficient at the luminescence energy \cite{blakemore1982}. For a circularly-polarized excitation, one also measures the profile of the difference between the $\sigma^+$- and   $\sigma^-$-polarized components of the luminescence. Its cross section is related to the spin density $s=n_+-n_-$ where  $n_\pm$ are the concentrations of electrons of spin $\pm$, taking the $z$ axis for quantization of the electronic spins, and is given by   
\begin{equation} \label{imagediff} I_d(r) = - A\int_0^d s(r,z)\exp\left[-\alpha_lz\right] \mathrm{d}z \end{equation} 
Finally, the  profile of the electronic spin polarization $\mathscr{P}(r) =s/n$ is given by 
\begin{equation} \label{imagepol}  I_d(r)/I_s(r)=\mathscr{P}(r) \mathscr{P}_i.
 \end{equation}
where  $\mathscr{P}_i=(g_+-g_-)/(g_++g_-)$ such that  $g_{\pm}$ is the spatially-dependent rate of creation of electrons of spin $\pm$. The quantity $g_{\pm}$ depends on the matrix elements of the allowed optical transitions and is equal to $\pm 0.5$ for $\sigma^{\mp}$ light excitation \cite{meier1984}. \

The densities $n(r, z)$ and $s(r, z)$ are, respectively, solutions of the continuity equations
\begin{equation}
(g_+ + g_-) -n/\tau +\frac{1}{q}\vec \nabla \cdot \left(\vec J_c \right)  = 0
\label{J2}
\end{equation}	 				
	\begin{equation}
(g_+ - g_-) -s/\tau_s+\frac{1}{q}\vec \nabla \cdot \left(\vec J_s \right )  = 0
\label{J3}
\end{equation}	 
Here $q$ is the absolute value of the electronic charge, $1/\tau = K_r(N_A + \delta p)$  and $1/\tau_s = 1/\tau + 1/T_1$, where $N_A$ is the acceptor density, $\delta p$ is the density of photocreated holes and $K_r$ is the bimolecular recombination coefficient. Since the spin relaxation time $T_1$ is long with respect to the various times which characterize spin transport, one considers separately the currents $\vec J_+$ and $\vec J_-$  of $+$ and $-$ spins so that, in a simple picture at low density,  $\vec J_c= \vec J_+ + \vec J_-= q D \vec \nabla n$ and $\vec J_s= \vec J_+ - \vec J_-= qD_s \vec \nabla s$ where $D $ and $D_s$ are the charge and spin diffusion constants.  These equations are solved by imposing i)  electron currents at the front ($z=0$) and back surface ($z=d$) that are equal to $qSn(0)$  and $-qS'n(d)$ respectively. Here $S$ and $S'$ are the corresponding  recombination velocities. ii) Spin currents equal to $qSs(0)$ and $-qS's(d)$, respectively. One can then define an effective lifetime $\tau_{eff}$ which takes account bulk and surface recombination and an effective spin lifetime $\tau_{seff}$ such that $1/\tau_{seff} = 1/\tau_{eff} + 1/T_1$ \cite{cadiz2013}. Charge and spin effective diffusion lengths are defined as $L_e^{eff}= \sqrt{D\tau_{eff}}$ and $L_s^{eff}= \sqrt{D_s \tau_{seff}}$. \

Figure \ref{figexp} also shows the sum and difference images at 15 K at a very low excitation power of 1.5 $\mu$W, as well as their angular-integrated cross sections. As seen from the  cross section of the laser profile, the sum and difference signals are observed well beyond the laser spot, and reveal charge and spin diffusion of photoelectrons after creation, respectively. Analysis of these profiles, using Eq.  (\ref{imagesum}), Eq. (\ref{imagediff}), Eq.  (\ref{J2}) and Eq. (\ref{J3}), gives  $L_e^{eff}=1.42$ $\mu m$ and $L_s^{seff}=1.25$ $\mu m$ at $T=15$K and shows that the spin relaxation time, $T_1$, is larger  than the electron lifetime.

\begin{figure*}[tbp]
\includegraphics[width=18cm, trim=1 1 1 500, clip=true] {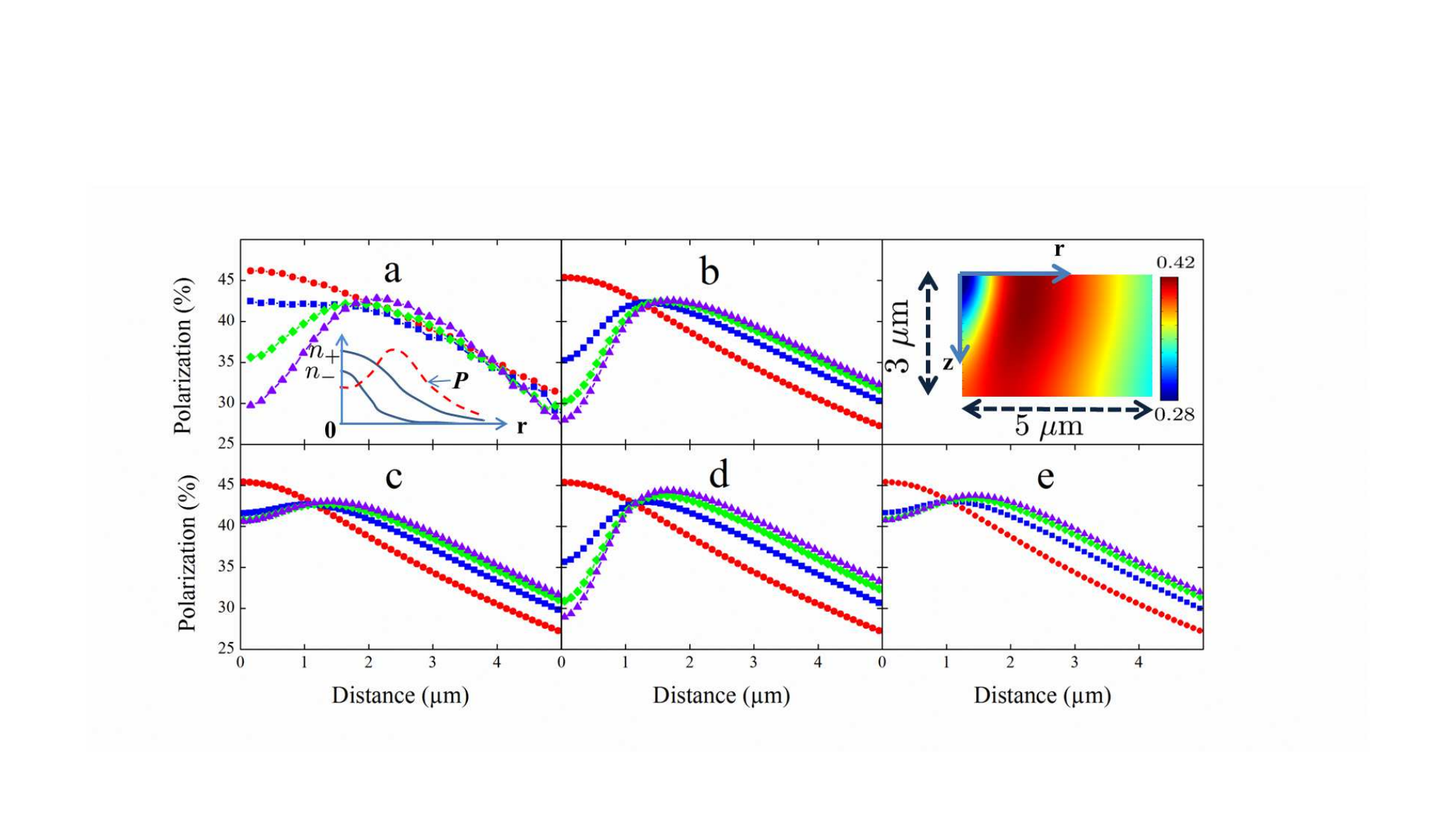}
\caption{Panel a shows the experimental polarization profiles at $T= 15\; K$ as a function of excitation power :  $28$ $\mu$W (filled circles), 1.03 mW (filled squares), 1.89 mW (filled diamonds), $2.55$ mW (filled triangles). The inset of Panel a interprets the formation of a polarization dip, as caused by  the larger diffusion length of majority electrons, which induces a depletion of these electrons at $r=0$. Panel b shows the  corresponding calculated profiles including all effects which modify spin transport. Panel c shows the image of the calculated spatial distribution of the polarization for an excitation power of $2.55$ mW. With respect to Panel b, the profiles of Panel d do not consider  ambipolar diffusion, those of Panel e do not consider thermoelectric currents and those of Panel f neglect both ambipolar diffusion and temperature gradients and thus illustrate the conditions of spin-grating experiments.}
\label{fig05}
\end{figure*}

\subsection{Experimental investigation of spin transport}

\subsubsection{Polarization profiles as a function of power}
 Panel a of Fig. \ref{fig05} shows the profiles of the electronic polarization $\mathscr{P}$,  at $T=15$K for increasing excitation powers, as obtained using  Eq.  (\ref{imagepol}) (the other panels are calculations to be explained in Sec. IV below). Curve a, taken at the same low power as Fig. \ref{figexp}, reveals the expected polarization decrease caused by spin-lattice relaxation during transport \cite{favorskiy2010}. Note that the low power electronic polarization at $r = 0$, $\mathscr{P}^{lp} (0)=45\;\%$ is almost equal to the initial polarization $ \mathscr{P}_i$. Since as will be shown below the spin relaxation time of thermalized electrons is much larger  than their lifetime at the excitation spot, the slight difference is attributed to spin-lattice relaxation during thermalization. As a result, in Eq.  (\ref{J2}) and Eq.  (\ref{J3}), $g_{+}$ and $g_{-}$ must be replaced by $g_{+}^*$ and $g_{-}^{*}$, respectively  such that $g_{+}^{*}(0)/g_{-}^{*}(0)=(1+\mathscr{P}^{lp} (0))/(1-\mathscr{P}^{lp}(0))$.  \
 
 As the power is increased, a polarization dip at $r=0$ progressively appears. At $2.55$ mW, the polarization at $r=0$ is $28\;\%$, while at $r\approx 2\; \mu m$, it is $42 \; \%$, slightly larger than its low power value at the same distance from the excitation spot. If the power is further increased, the profiles, not shown here, exhibit an overall decrease of the polarization because of heating of the electron gas by the laser.\
 
In agreement with the dependences of the polarization dip as a function of excitation power and temperature, it has been shown \cite{cadiz_prl2013} that this effect occurs because of Pauli blockade in the degenerate photo-electron gas, i. e.  when either one or both of $n_+$ and $n_-$ become larger than the spin-resolved effective density of states in the conduction band $N_{c}^s$, given by 
\begin {equation}
N_{c}^s= (1/2)N_{c}^0(T_e/300)^{3/2} 
\label{Nc}
\end{equation}
where $N_{c}^0=4.7 \times 10^{17}$ cm$^{-3}$ and $T_e$ is the temperature of the electron gas. In this case, the diffusion constant is larger for majority spin electrons ($D_+$) than for minority spin electrons ($D_-$). The more efficient removal by diffusion away from $r=0$  induces a depletion of majority electrons at $r=0$, with a relative accumulation at some distance away from $r=0$. This effect is illustrated in the inset of Fig. \ref{fig05}(a)  which shows the spatial dependences of $n_+$ and $n_-$. In this framework, the ratio of $D_+/D_-$ can be estimated  by considering a simple two-dimensional picture, where the concentrations $n_{\pm}$ are replaced by their averages $<n_{\pm}>$ over $z$, and by writing that the diffusion time out of the excitation spot is, within numerical factors of order unity, given by 
\begin{equation}
\label{taueff0}
\tau_{eff}^{0} (<n_{\pm}>)  \approx \omega^{2}/[4D_{\pm})] 
\end{equation}
and is of the order of several ps that is, shorter than characteristic times for  recombination and spin relaxation. Considering that diffusion is the dominant process for removal of electrons from the excitation spot, the spin concentrations at $r=0$ are given by 
\begin{equation}
\label{eqcalcnpm}
 <n_{\pm}> \approx g^{*}_{\pm}(0) \tau_{eff}^{0} (<n_{\pm}>)
\end{equation}
One then obtains the following very simple result, in which the poorly known numerical factors of Eq. (\ref{eqcalcnpm}) are eliminated 
\begin{equation}
\label{ratio}
D_{+} /D_{-} = \frac{1+\mathscr{P}^{lp}(0)}{1-\mathscr{P}^{lp}(0)}\times \frac{1-\mathscr{P}(0)}{1+\mathscr{P}(0)} \end{equation} 

\noindent
At high power, we find  $D_{+} /D_{-}  \approx 1.49$ implying that degeneracy causes a significant spin-dependence of the diffusion constant. Writing to first order
	\begin{equation}
\label{eta}
   D_{\pm} = D^*[ 1 \pm \delta \mathscr{P}]
    \end{equation} 
where the expressions for $D^*$ and $\delta$ will be given below, one finds $\delta= 0.65$.

\begin{figure}[tbp]
\includegraphics[width=8 cm, trim=1 1 1 300, clip=true] {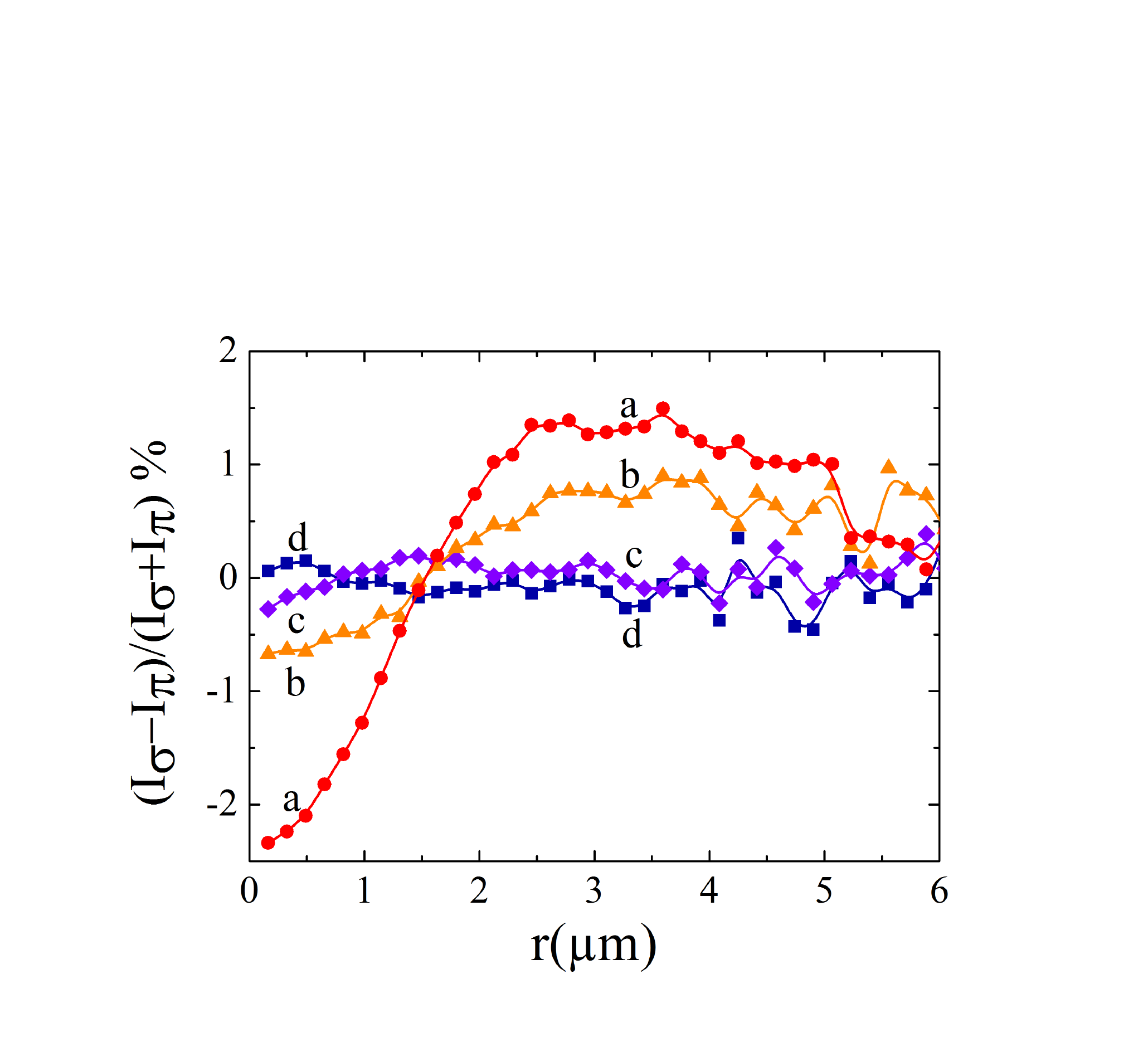}
\caption{Relative difference between the luminescence intensity profiles obtained under circularly-polarized excitation ($\sigma$) and under linearly-polarized excitation ($\pi$), for different excitation powers, a) 2.33 mW, b) 0.95 mW, c) 0.41 mW and d) 65 nW. For each excitation power, the only difference is the polarization of the photoelectron gas.
A difference of the order of $2.5 \;\%$ between both profiles is observed at high power at $r=0$, revealing the spin-dependent diffusion of photoelectrons. }
\label{pisigma}
\end{figure}

\subsubsection{Spin-dependent charge diffusion.}
The effective charge diffusion constant, defined as $\langle D \rangle= (1/n ) \sum_i n_i D_i$, is found using Eq.  (\ref{eta}) and given by 
\begin{equation}   
 \langle D \rangle  =D^*[1+ \delta \mathscr{P}^2]
 \label{Qpressure}
 \end{equation} 
 \noindent 
which implies that the sum profile under degeneracy depends on spin via a second order effect. In order to show such effect, the sum profiles  $I_{\sigma}$ for a circularly-polarized ($\sigma$) excitation were compared with the profiles $I_{\pi}$ for a linearly-polarized excitation, ($\pi$, so that $\mathscr{P}=0$), keeping the excitation power constant to within $0.1 \%$. Figure \ref{pisigma} shows the relative difference of these profiles at $T=15\;K$ for different power densities. At low power (Curve d), the signal is zero within experimental uncertainty, showing that charge transport  in nondegenerate conditions does not depend on spin. In contrast, when the excitation power is increased, there progressively appears a depletion of photoelectrons at $r=0$. This depletion, of the order of $2.5 \;\%$, is compensated by a converse excess of photoelectrons at a distance larger than about 1.5 $\mu$m. This shows that the diffusion constant of spin-polarized electrons is larger than for spin-unpolarized electrons created by $\pi$ excitation.\

Using  Eq. (\ref{Qpressure}), the spin-dependence of the charge concentration at $r=0$  is given by 
\begin{equation}
\frac{<n_{\sigma}>-<n_{\pi}>}{<n_{\sigma}> +< n_{\pi}>} = \frac{\langle D_{\pi} \rangle - \langle D_{\sigma}\rangle}{\langle D_{\pi} \rangle + \langle D_{\sigma}\rangle} = -\frac{ \delta \mathscr{P}^2}{2 +  \delta \mathscr{P}^2}
\end{equation}

\noindent
from which we obtain, in agreement with the preceding subsection, $\delta= 0.58$.

\begin{figure}[tbp]
\includegraphics[width=8 cm, trim=1 1 1 300, clip=true] {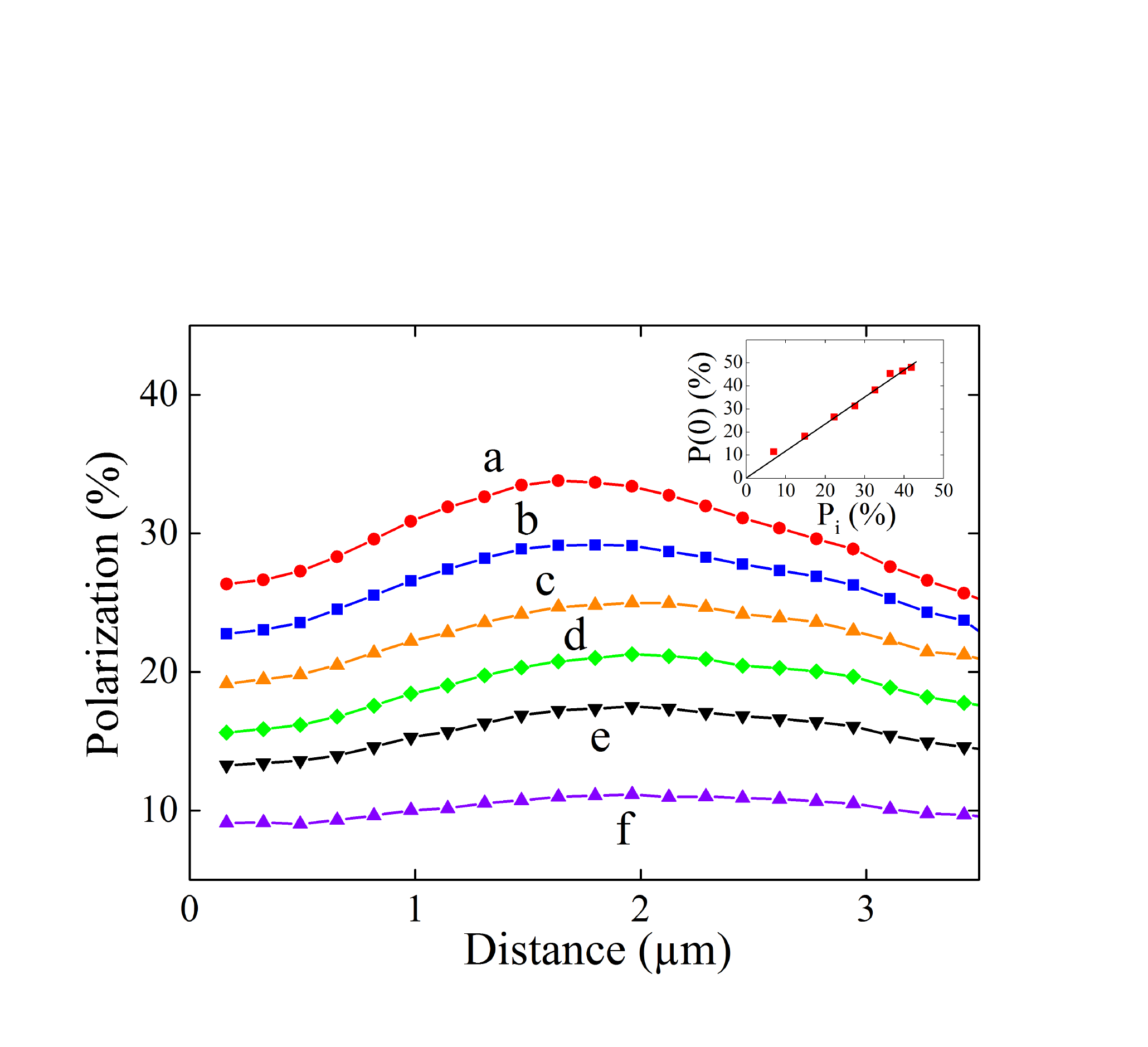}
\caption{Polarization profiles for decreasing values of the initial polarization $ \left|\mathscr{P}_i  \right|$, equal to 0.5 (a), 0.43 (b), 0.39 (c), 0.33 (d), 0.27 (e) and 0.18 (f). As shown in the inset, the electronic polarization at $r=0$ is proportional to  $\mathscr{P}_i$, thus revealing that the ratio $D_+/D_-$ depends linearly on electronic polarization. }
\label{Figpi}
\end{figure}
\subsubsection{Polarization profiles as a function of excitation light polarization}

In order to investigate the dependence of the effect of Pauli blockade on electronic polarization, the helicity of the excitation light is changed in order to change $\mathscr{P}_i$. The corresponding polarization profiles, shown in Fig. \ref{Figpi},  show that the dip at $r=0$ indeed decreases with decreasing  $\mathscr{P}_i$. Furthermore, as shown in the inset of Fig. \ref{Figpi}, the electronic polarization  at $r=0$ is proportional to $\mathscr{P}_i$. This behavior is in agreement with the  predictions made using Eq.  (\ref{ratio}) and Eq.  (\ref{eta}), according to which,  to first order in  $\mathscr{P}$, one has $\mathscr{P}(0) = \mathscr{P}^{lp} (0)/(1+ \delta)$. From the slope of this behavior, one finds  $\delta= 0.45$, in qualitative agreement with the value of the preceding subsection.\ 

\subsubsection{Kinetic energy effects}
\begin{figure}[tbp]
\includegraphics[width=8 cm, trim=1 1 1 400, clip=true] {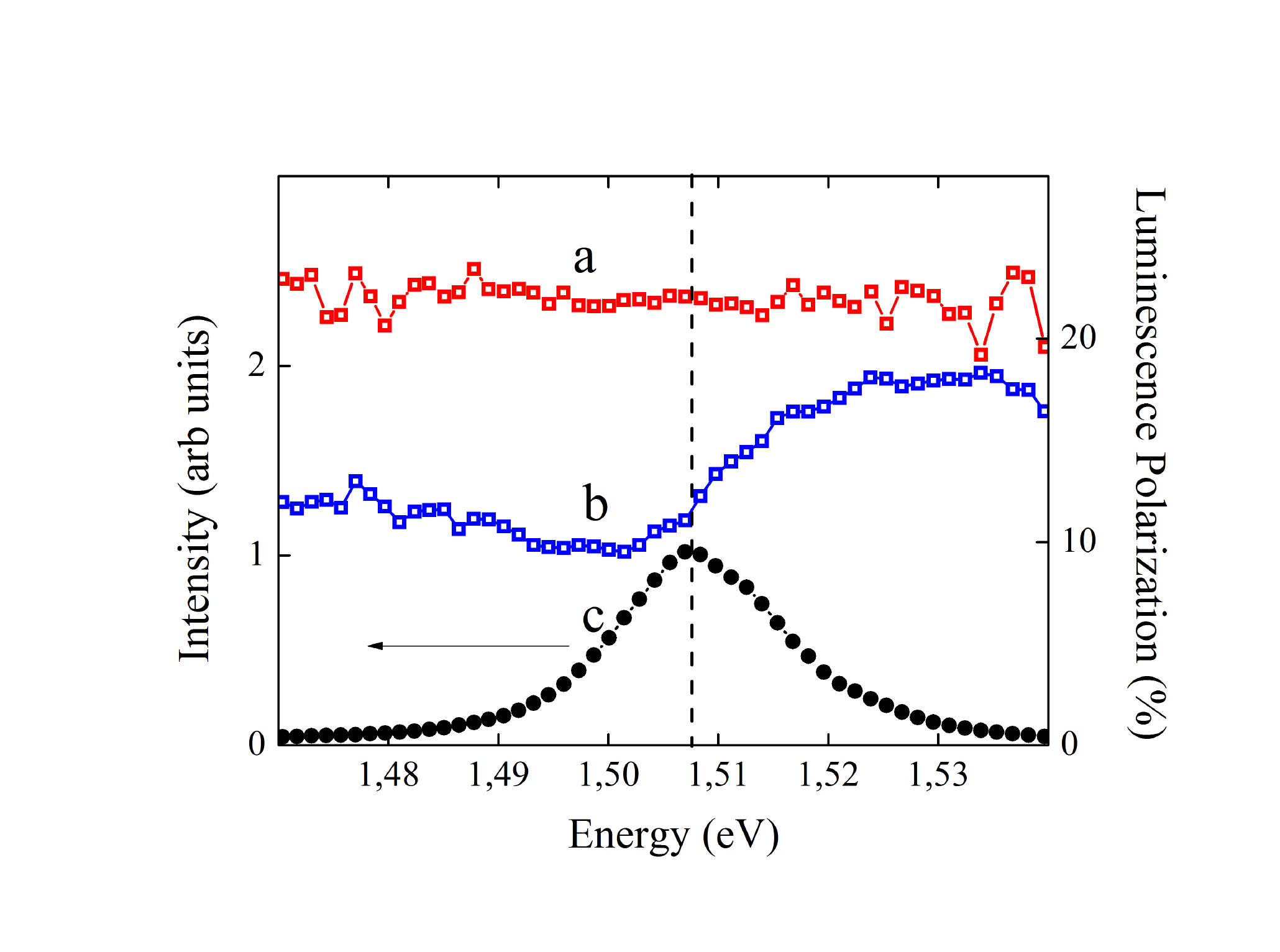}
\caption{Dependence of spin polarization on kinetic energy. Curve a and Curve b show the spatially-resolved polarization spectrum at $r=0$  and at 15 K for a power of 28 $\mu$W and 2.55 mW, respectively. Curve c shows for comparison the intensity spectrum at 2.55 mW at the place of excitation. Comparison between Curves a and b shows that the Pauli blockade effect is smaller for hot electrons. }
\label{figsp}
\end{figure}

 The luminescence and polarization spectra can be monitored as a function of $r$ by using a scanned multimode optical fiber that captures PL over a spot size of $0.9\;\mu m$. The fiber is then coupled to a spectrometer to yield a local spectrum like those shown in Fig. \ref{figsp}. The fact that the luminescence lies at an energy  smaller than the GaAs bandgap ($1.519$ eV at this temperature) has been interpreted as due to bandgap renormalisation caused by the free hole population Ref. \cite{feng1995,casey1976}. In nondegenerate  conditions (Curve a), the polarization does not depend on light energy and is consistent with the electronic polarization at $r=0$  in Panel a of Fig. \ref{fig05}. As expected, in degenerate conditions (Curve b), the overall polarization  is weaker than for Curve a because of the spin-dependent transport effects discussed above. However, this polarization decrease is mostly observed on the low energy side of the spectrum, while for energies above 1.52 eV, the two spectra almost coincide. It is concluded that the spin filter effect decreases with increasing kinetic energy in the conduction band.\

As suggested in Ref. \cite{Volkl2011} in the case of [110] quantum wells, it is tempting to conclude that the depolarization of thermalized electrons at $r=0$ rather arises from an increased efficiency of the local spin relaxation processes, caused by the larger hole concentration or by the increased temperature.  This hypothesis cannot explain the results for three main reasons : i) Such polarization loss can only concern electrons localized in potential fluctuations, since diffusive electrons will transmit their depolarization after diffusion. However,  localized electrons only appear at lattice temperatures smaller than 10K  and are absent at the present higher temperature \cite{cadiz2014}. ii) Since the effective lifetime at $r=0$ is $\omega ^2 /4D  \approx 10$ ps, the polarization decrease would require an extremely strong, unphysical, decrease of $T_1$ from its value of $1125$ $ps$ at low power \cite{cadiz2014}.  iii) Since an increased spin relaxation at $r=0$ does not affect the   charge, the present hypothesis cannot explain the observed dependences of the charge diffusion on intensity and polarization reported in Fig. \ref{pisigma}.

\section{Theory}

\subsection{Charge and spin unipolar diffusion equations}

The charge and spin currents $\vec J_{c}$ and $\vec J_{s}$ which appear in the diffusion equations [Eq.  (\ref{J2}) and Eq. (\ref{J3})] are expressed as the  sum of contributions of diffusive charge and spin currents, of  drift currents due to internal electric fields of ambipolar origin, and of thermoelectric  currents caused by local heating of the photoelectron gas : $\vec J_{c(s)}= \vec J_{c(s)}^{dr}+ \vec J_{c(s)}^{dif}+\vec J_{c(s)}^{T}$  where each of these contributions is the sum and difference of the corresponding spin currents, respectively. These currents are calculated below. 
 
\subsubsection{Drift currents}

The drift current of electrons of spin $i$ is given by $\vec J_{i}^{dr}=\vec E \sum_j \sigma_{ij}$  where the nondiagonal elements of the conductivity matrix $\sigma_{ij}$ reflect the coupling  between opposite spins (spin Coulomb drag), mostly originating from electron-electron collisions \cite{weber2005, takahashi2008, vignale2002}. It is given by 
\begin{equation}
\sigma_{ij}  =qn_i\mu_i \alpha_{ij}
\label{sigma}
\end{equation}
\noindent
where the  mobility $\mu_i$  of electrons of spin $i$ is given  by 
\begin{equation}
\mu_i = q\tau_{mi}/m^*
\label{eqmu}
\end{equation} 

The momentum relaxation time $\tau_{mi}$, calculated in Appendix A using the Boltzmann equation formalism, is equal to   
 \begin{equation}
\tau_{mi} =-\frac{2}{3}.\frac{\int \tau_{m}(\varepsilon) \varepsilon^{3/2} (\partial f_{0i}/\partial \varepsilon) d\varepsilon}{\int \varepsilon^{1/2} f_{0i}d\varepsilon} 
\label{taumi}
 \end{equation} 
\noindent
where $f_{0i}$ is the Fermi distribution and $\varepsilon$ is the kinetic energy.
Here, $\tau_{m}(\varepsilon)$ is assumed to be of the form \cite{smith1978, flatte2006}
 \begin{equation}
\tau_m(\varepsilon) \propto \varepsilon^{p}. 
\label{s}
 \end{equation}
where $p$ depends on the scattering process which determines the mobility. The concentration dependence of $\mu_i$ is obtained using  Eq. (\ref{eqmu}) and Eq. (\ref{taumi}).  One finds
\begin{equation} 
\mu_{i} = \mu_0 \zeta(n_i)=  \mu_0 \frac{\mathscr{F}_{p+1/2}^*(\eta_{i})}{\mathscr{F}_{1/2}^*(\eta_{i})}
\label{mu}
\end{equation}
\noindent
where $\mu_0$ is the mobility in nondegenerate conditions, $\eta_i=E_{Fi}/k_BT_e$ where $E_{Fi}$ is the Fermi energy and $k_B$ is the Boltzmann constant,  and where the Fermi integral $\mathscr{F}^{*}_{k}(\eta_i)$ is given by 
\begin{equation} 
\mathscr{F}_k^*(\eta_i)= \frac{1}{\Gamma (k+1)}\int_{0}^{\infty} \frac {x^{k}dx}{1+exp(x-\eta_i)} 
\label{fermiint}  
\end{equation} 
Here $\eta_i$  is related to the electronic spin concentration by
\begin{equation}  
 n_i= N_{c}^s \mathscr{F}_{1/2}^*(\eta_i)
 \label{conc} 
\end{equation}     
 \noindent
The coefficients $\alpha_{ij}$, given by \cite{vignale2002}
\begin{equation}
\alpha_{ij} = \frac{\tau_{ee}\delta_{ij} + (n_j/n)\tau_{m, -i}}{\tau_{ee}+\tau_m} 
 \label{eqE1}\end{equation} 
\noindent
account for the conductivity changes of each spin reservoir caused by the spin-spin couplings. Here $\delta_{ij}$ is the Kronecker symbol. The time $\tau_{ee}$ is given by $n\tau_{ee} = \tau_{ee,i}n_{-i}$, where $\tau_{eei}$ is  the collision time for an electron with spin $i$ with an electron of opposite spin.  The spin-averaged time $\tau_m$ is given by 	  			
\begin{equation}\tau_m =  (n_+/n)\tau_{m_-} +  (n_-/n)\tau_{m_+} \label{E2}\end{equation}

One has finally  $\vec J_{c}^{dr}=\sigma_c \vec E$ and $\vec J_{s}^{dr}=\sigma_s \vec E$, where  $\sigma_c = \sum_{ij}\sigma_{ij}$ and $\sigma_s= \sum_{ij} i\sigma_{ij}$. 
  
\subsubsection{Diffusive currents}
 
 The diffusive current of electrons of spin $i$ depends on the spatial gradients of the Fermi energies $E_{Fj}$, and is given by $q\vec J_{i}^{dif}=  \sum_j\sigma_{ij}( \vec \nabla_r E_{F_j}\mid_{T_e})$ where the spatial gradient of the Fermi energy at constant  temperature of the electron gas $T_e$ is expressed as  $\vec \nabla_r E_{F_j}\mid_T= \sum_i S_{ji} \vec \nabla_r n_i $ where the spin stiffness matrix is given by
\begin{equation}
S_{ij}= \frac{\partial E_{F_i}}{\partial n_j}.
\label{kimatrix}
\end{equation}  
Note that the total Fermi level gradient is given by $\vec \nabla_r E_{F_j}=\vec \nabla_r E_{F_j}\mid_{T_e} + (\partial E_{F_j}/\partial k_B T_e)\vec \nabla_r k_B T_e$. However, the second term contributes to the thermoelectric current and will  be considered in the following subsection. The diffusive current  can be rewritten as 
\begin{equation}
\ \vec J_{i}^{dif}= q(D_{ii} \vec \nabla n_i + D_{i, -i} \vec \nabla n_{-i})
\label{F}
\end{equation}
where the elements of the diffusion matrix $D$ are given by 
\begin{equation}
qD_{ij}= \alpha_{ii} n_i \mu_i S_{ij} +\alpha_{i, -i} n_{-i}\mu_{-i} S_{-i, j}.
\label{Dmatrix}
\end{equation}
\noindent

Equation (\ref{Dmatrix}) is the generalized Einstein relation. The charge and spin diffusive currents are finally given by 
\begin{equation}
\frac{1}{q} J_c^{dif}= D_{cc} \vec \nabla n + D_{cs} \vec \nabla s
\label{difcharge}
\end{equation}
\noindent

\begin{equation}
\frac{1}{q} J_s^{dif}= D_{sc} \vec \nabla n + D_{ss} \vec \nabla s
\label{difspin}
\end{equation}
\noindent
where the diffusion constants are linear combinations of the $D_{ij}$ given by $2D_{cc} = \sum_{ij}D_{ij}$ and $2D_{ss} = \sum_{ij}ij D_{ij}$, $2D_{cs} = \sum_{ij} j D_{ij}$ and $2D_{sc} = \sum_{ij}i D_{ij}$, and can be straightforwardly calculated if the spin stiffness matrix $S_{ij}$ is known.  It is concluded that two types of spin-related mechanisms can affect charge and spin diffusion. Spin-spin couplings result in a non diagonal $D_{ij}$ matrix, because of which a gradient of spins $j$ affects the diffusive current of spins $i$. Charge-spin and spin-charge couplings originate from non zero values of  $D_{cs}$ and $D_{sc}$, respectively and result in a dependence on the spin(charge) current on the charge (spin) density. 

\subsubsection{Thermoelectric currents : Soret charge and spin currents} 

The thermoelectric current of electrons of spin $i$ is calculated in Appendix A by solving the Boltzmann equation. It is of the form  $\vec J_{i}^T=-\sum_j \sigma_{ij} \mathscr{S}_{j} \vec \nabla_r T_e$ and $\mathscr{S}_{j}$ is the spin-dependent Seebeck coefficient  for which the value for unpolarized electrons is equal to its usual value given elsewhere \cite{CutlerMott1969}. It is given by  $\mathscr{S}_{j}=-(1/qT_e)(E_{Tj}-\gamma_j k_B T_e)$, where   

\begin{equation}
	 E_{Ti} = \frac{\int \tau_{m}(\varepsilon) \varepsilon^{3/2} (\partial f_{0i}/\partial \varepsilon) d\varepsilon}{\int \tau_{m}(\varepsilon) \varepsilon^{1/2} (\partial f_{0i}/\partial \varepsilon) d\varepsilon} 
	\label{thermalen}
	\end{equation}
\noindent
where $\gamma_i$ depends on the Fermi integral $\mathscr{F}_k(\eta) = \Gamma (k+1)\mathscr{F}_k^*(\eta)$ and  is given by 
  
  \begin{equation}
\gamma_i=\frac{\mathscr{F}_{1/2}(\eta_i)}{\mathscr{F}_{-1/2}(\eta_i)}
\label{gamma}
\end{equation} 
  
The above equations are similar to those of the Seebeck effect in  which there is however no current \cite{brechet2010}. Here the spin currents arise through a distinct effect, which has been described by Charles Soret \cite{soret1879} for mass transport. The current $\vec J_{i}^T$ will be hereafter called the Soret current. Here, it is more convenient to express it in the form   
\begin{equation}
	\vec J_{i}^T=q \sum_j \vec K_{ij}n_j  
	\label{thermalcurrent}
	\end{equation}	
\noindent
The Soret velocity matrix $\vec K_{ij}$ is given by 
	\begin{equation}
	q\vec K_{ij}= \alpha_{ij}\mu_{j}(\frac{E_{Tj}}{k_B T_e}-\gamma_j)\vec \nabla_r (k_B T_e)
	\label{matrixK}
	\end{equation}
\noindent

The currents $\vec J_{cT}$ and $\vec J_{sT}$ defined in Eq. (\ref{J2}) and Eq. (\ref{J3}) are finally given by 
\begin{equation}
\frac{1}{q} \vec J_{c}^T =\vec K_{cc} n + \vec K_{cs} s 
\label{thermalcharge}
\end{equation}
 \begin{equation}
\frac{1}{q} \vec J_{s}^T  = \vec K_{sc} n + \vec K_{ss} s 
\label{thermalspin}
\end{equation}
\noindent
where $2\vec K_{cc} = \sum_{ij} \vec K_{ij}$, $2\vec K_{ss} = \sum_{ij} ij \vec K_{ij}$, $2\vec K_{sc} = \sum_{ij} i \vec K_{ij}$, and $2\vec K_{cs} = \sum_{ij} j \vec K_{ij}$.\

\subsection{Ambipolar diffusion equations} 

Taking account of all contributions defined in the preceding section, the diffusion equations for electrons and spins can finally be written 

\begin{equation}    (g_++g_-)-n/\tau + \vec \nabla \cdot [(\vec E/q)\sigma_c + D_{cc} \vec \nabla n + D_{cs} \vec \nabla s + \vec J_c^T]=  0
 \label{electrons} \end{equation}

\begin{equation}    (g_+-g_-)-s/\tau _s + \vec \nabla \cdot [(\vec E/q)\sigma_s + D_{sc} \vec \nabla n + D_{ss} \vec \nabla s + \vec J_s^T]=  0
 \label{spins} \end{equation}
 
In order to take account of the electrostatic coupling between electrons and the slower diffusing holes, it is further necessary  to couple these equations with the diffusion equation for spin-unpolarized holes, which is  
\begin{equation}    (g_++g_-)-\delta p/\tau + \vec \nabla \cdot [-(\vec E/q)\sigma_h + D_h \vec \nabla \delta p ]=  0
 \label{holes} \end{equation}
\noindent
where $\delta p$ is the photohole concentration and $D_h$  is the hole diffusion constant. Here $\sigma_h= q(N_A^- +\delta p)\mu_h$ is the  conductivity, where $\mu_h$ is the hole mobility. The thermoelectric hole current  is neglected since the local heating of the hole gas is weak \cite{leo1991}. The electric field satisfies Poisson's equation 
\begin{equation}    
\vec \nabla \cdot \vec E= \frac{e}{\epsilon \epsilon _0}  (\delta p - n)
\label{Poisson} 
\end{equation}
where $\epsilon$ is the dielectric constant and $\epsilon_0$ is the permittivity of free space. Equations (\ref{electrons}),  (\ref{spins}), (\ref{holes}) and (\ref{Poisson})  are solved numerically, as shown in Appendix B, by imposing that, in addition to the boundary conditions for Eq. (\ref{J2}), (\ref{J3}) defined in Sec. IIA, the hole currents  at the front ($z=0$) and back surface ($z=d$) are equal to $qS\delta p(0)$  and $-qS'\delta p(d)$. 

\section{Relevant mechanisms for spin transport under local excitation of $p^+$ material}

\subsection{Effect of degeneracy on diffusion and its spin-dependence}
 As already shown in Ref. \cite{cadiz_prl2013}, degeneracy can induce a spin dependence of the diffusion constant due to two distinct effects which are direct consequences of the Pauli Principle. The first one is the concentration dependence of the spin stiffness  [Eq.  (\ref{kimatrix})]. Neglecting  electron-electron interactions  which will be shown below to be screened by the hole gas,  the spin stiffness matrix is diagonal and, in Eq. (\ref{sigma}),  $\alpha_{ij}=\delta_{ij}$. Using Eq. (\ref{conc}), one finds $S_{ii}=k_BT_e/[\mathscr{F}_{-1/2}^*(\eta_i)N_c^s]$ and  Eq. (\ref{Dmatrix})  reduces to the spin-uncoupled Einstein equation for a degenerate electron gas \cite{finkelshtein1983} 
\begin{equation}
D_{i} = \frac{n_i \mu_i}{q} S_{i i} =\xi(n_i)\mu_i \frac{k_BT}{q}
\label{D}
\end{equation}
where $\xi= n_i S_{ii}/k_BT_e$ is given by  
\begin{equation}
 \xi(n_i)=\frac{\mathscr{F}_{1/2}^*(\eta_i)}{\mathscr{F}_{-1/2}^*(\eta_i)}= 2 \gamma_i
 \label{xi}
\end{equation}
This quantity is unity for a nondegenerate gas and increases with concentration. \

The second possible effect induced by degeneracy is a spin-dependent increase of the mobility, as described by  Eq. (\ref{mu}) and is a direct consequence of Pauli exclusion due to which elementary scattering processes are forbidden  if the final state is already occupied by an electron of the same spin. The diffusion constants for spins $\pm$ are finally given by 
\begin{equation}
D_i =   D_0 \nu(n_i) \end{equation}
\noindent
 where $D_0=\mu_0 k_B T_e/q$  and \begin{equation}
 \nu(n_i)= \xi(n_i) \zeta(n_i)= \frac{\mathscr{F}_{p+1/2}^* ( \eta_{i} )}{\mathscr{F}_{-1/2}^*(\eta_{i})}.
 \label{nu}
 \end{equation}\
 
 
\noindent
Using the linearized form defined by Eq. (\ref{eta}), the charge and spin diffusion constants which appear in Eq.  (\ref{electrons}) and Eq.  (\ref{spins}) are finally given 
\begin{equation}    
D_{cc}=D_{ss}=D_0\nu(n/2) 
\label{K1} 
\end{equation}
\begin{equation}    
D_{cs}=D_{sc}=D_0 \nu(n/2)\delta \mathscr{P} 
\label{L1} 
\end{equation}
\noindent
where the two quantities $\delta$ and $D^*$ introduced in Eq.  (\ref{eta}) are now given a precise definition. Here $\delta= d\log [\nu(n/2)] /d\log (n/2)$ and is equal to  $ 2(p+1)/3$ at large degeneracy, while $D^*=D_0 \nu(n/2)$. The Pauli Principle induces a coupling between the charge and spin diffusions, for which the coupling coefficients $\delta\mathscr{P}$ are identical in the two equations. They increase with electron polarization and concentration.\

\begin{figure}[tbp]
\includegraphics[width=8 cm, trim=1 1 1 200, clip=true] {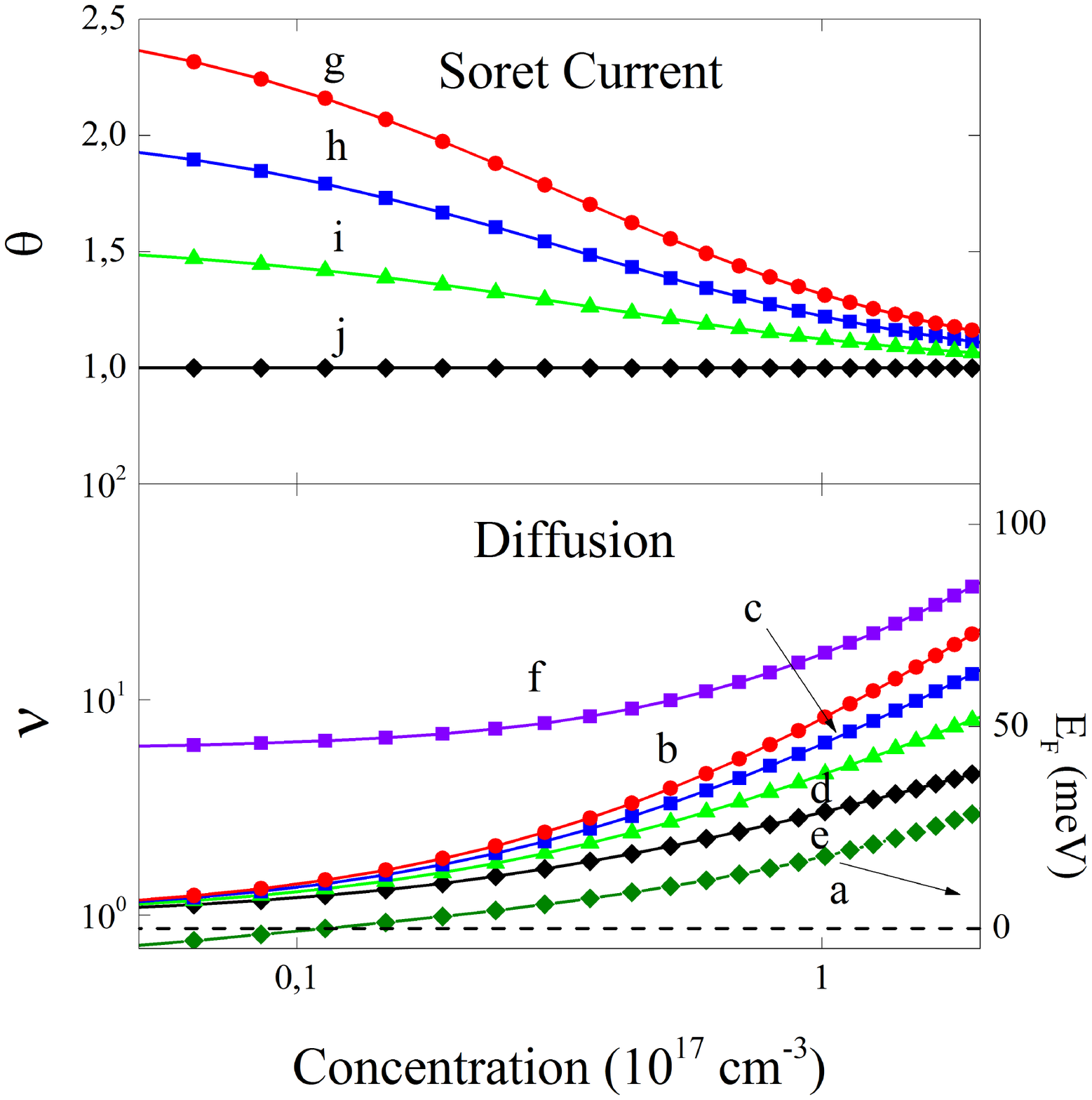}
\caption{The bottom panel shows, at $T_e=50$ K, the Fermi energy as a function of the spin-resolved photoelectron concentration $n_i$ (Curve a) revealing the onset of degeneracy near $n=10^{16}\;\mbox{cm}^{-3}$. The  dependence of the reduced diffusion constant $\nu$ [Eq. (\ref{nu})] is shown for selected values of $p$, as defined by Eq. (\ref{s}) :  $3/2$ (b), $1$ (c), $1/2$ (d), $0$ (e). Curve f shows this same quantity for $p=3/2$, but for hot electrons, of kinetic energy larger than $k_B T_e$. The top panel shows the quantity $\theta$ [Eq. (\ref{theta})] on which depends the Soret current, for $p=3/2$ (g), $1$ (h), $1/2$ (i), $0$ (j).}
\label{fig01}
\end{figure}

For highly p-doped GaAs and $T_e= 50K$, Curve a of Fig. \ref{fig01} shows the concentration dependence  of the Fermi energy $E_{F_i}$. Degeneracy is achieved for electronic concentrations larger than $10^{16}$cm$^{-3}$, for which $E_{Fi}>0$. As shown in the same panel, the reduced diffusion constant increases with concentration from  its values of $1$ in nondegenerate conditions. For $p=3/2$, it is found that the spin dependence of the spin stiffness and of the mobility have an equal importance in the spin dependence of the diffusion since the increase  of $\xi$ and $\zeta$, not shown in the figure, are quite similar. When $p$ is decreased, the concentration dependence of the mobility becomes reduced, which induces as shown in the figure, a reduction of the concentration and spin dependence of the diffusion constant. For a  typical electron gas of density $n_+ + n_-=10^{17}\;\mbox{cm}^{-3}$ and $p=3/2$, with a spin polarization of $\mathscr{P}= (n_+-n_-)/n=40\;\%$  one finds  $D_+/D_- \approx 2.2$. For $p=0$, this ratio is reduced to $D_+/D_- \approx 1.3$  for the same values of $n_{\pm}$. However, because of  the smaller diffusion constant with respect to $p=3/2$, the concentration at $r=0$ is increased in order to verify Eq.  (\ref{eqcalcnpm}), so that the effect of $p$ on the actual spin dependence of $D$ is rather weak.\

The above calculations can be extended to hot electrons in order to explain the results of Fig. \ref{figsp}. Using a restriction of the Fermi integrals appearing in Eq. (\ref{gamma}) and Eq. (\ref{mu}) to electrons of kinetic energy larger than  $\epsilon_{min}$ reduces the spin stiffness and the concentration dependence of $\xi$. This is the same for the mobility because the rate of occupation of electronic states at the corresponding kinetic energy is smaller, so that scattering by an ionized impurity is less likely to be forbidden by the Pauli principle. Curve f of Fig. \ref{fig01}, calculated using  $\epsilon_{min}= k_B T_e$  and in the particular case of $p=3/2$, shows that the concentration dependence of  $\nu$ is significantly reduced and explains the weaker spin dependence of the diffusion constant of hot electrons. \ 

\subsection{Spin Soret current under degeneracy}

Under the sole effect of the thermal gradient, the Soret velocities are given by $K_{cc}=K_{ss}=[K_{++}+K_{--}]/2$  and  $K_{cs}=K_{sc}=[K_{++}-K_{--}]/2$ while $K_{+-}=K_{-+}=0$. The ratio of the unipolar diffusive [$J_c^{dif}$] and Soret currents is then given by  
\begin{equation}
\frac {J_c^{T}}{J_c^{dif}}  = \theta \frac{\vec{\nabla}_r T_e/T_e}{\vec{\nabla}_r n/n}
\label{compsoret} 
\end{equation}	 
\noindent
The dimensionless quantity $ \theta$, given by 
\begin{equation}
\theta= \frac {E_T}{\xi k_B T} -1/2
\label{theta} 
\end{equation}	
is related to the Seebeck constant defined in Sec. III A3 by $\mathscr{S}=- (k_B/q)\xi \theta$ and is for spins $i$ given by 
\begin{equation}
2\theta(n_i)= \frac{p+3/2}{p+1/2}\frac{\mathscr{F}_{p+1/2} ( \eta_{i} )}{\mathscr{F}_{p-1/2}(\eta_{i})}\frac{\mathscr{F}_{-1/2} ( \eta_{i} )}{\mathscr{F}_{1/2}(\eta_{i})}-1
\label{thetaexp} 
\end{equation}
Shown in the top panel of Fig. \ref{fig01} are the concentration dependences of $\theta$ for selected values of $p$. Since $\theta$ is close to unity, the ratio of the Soret current to the usual diffusive current is mainly determined by the relative values of the temperature and charge gradients. While $\theta$ is unity for $p=0$, $\theta$ decreases for $p\neq 0$  with increasing concentration from $p+1$ in the nondegenerate limit to unity at very large $n_i$. \ 

As found from Eq. (\ref{matrixK}), 
\begin{equation}
\label{ratioK}
\frac{ K_{++}}{K_{--}}=\frac{\theta(n_+) }{\theta(n_-)} \frac{ D_+}{D_-}
\end{equation}
\noindent
so that under degeneracy the Soret current becomes spin-dependent in the same way as  diffusion. For $p\neq 0$, $\theta(n_+) < \theta(n_-) $, so that $K_{++}/K_{--}<D_+/D_-$. In this case, the thermal gradient causes an effective decrease of the polarization dip.  

\subsection{Hole screening of electron-electron interactions}
	 It is shown here that spin-spin or spin-charge couplings induced by electron-electron interactions are strongly reduced because of screening by the hole gas. The effect of hole screening can be simply taken into account in the Random Phase Approximation (RPA) in the present case where the hole screening is dominant over the electronic one. In this case, the static Coulomb potential in Fourier space is given by  \cite{mahan1981,collet1993} 
 \begin{equation}
 \label{potential}
 v(k)= \frac{4 \pi e^2}{\epsilon (k^2 +k_{DH}^2)} 
\end{equation}
where $e = q/\sqrt{4 \pi \epsilon_0}$. The Debye H\"uckel screening wave vector $k_{DH}$ depends on the hole concentration according to \cite{collet1986}
\begin{equation}
 \label{kappaDH}
 k^2_{DH}= \frac{4 \pi e^2 N_A}{\epsilon k_B T} \frac{1}{\xi (N_A+\delta p)}
\end{equation}
where the function $\xi(x)$ is related to the hole Fermi energy and defined by  Eq. (\ref{xi}). In this framework, it seems clear that electron-electron interactions will be decreased if $k_{DH}$ is larger than the typical value of $k$, of the order of the Fermi wavevector $k_{Fi} =(6\pi ^2  n_i)^{1/3}$.\

\begin{figure}[tbp]
\includegraphics[width=8 cm, trim=1 1 1 150, clip=true] {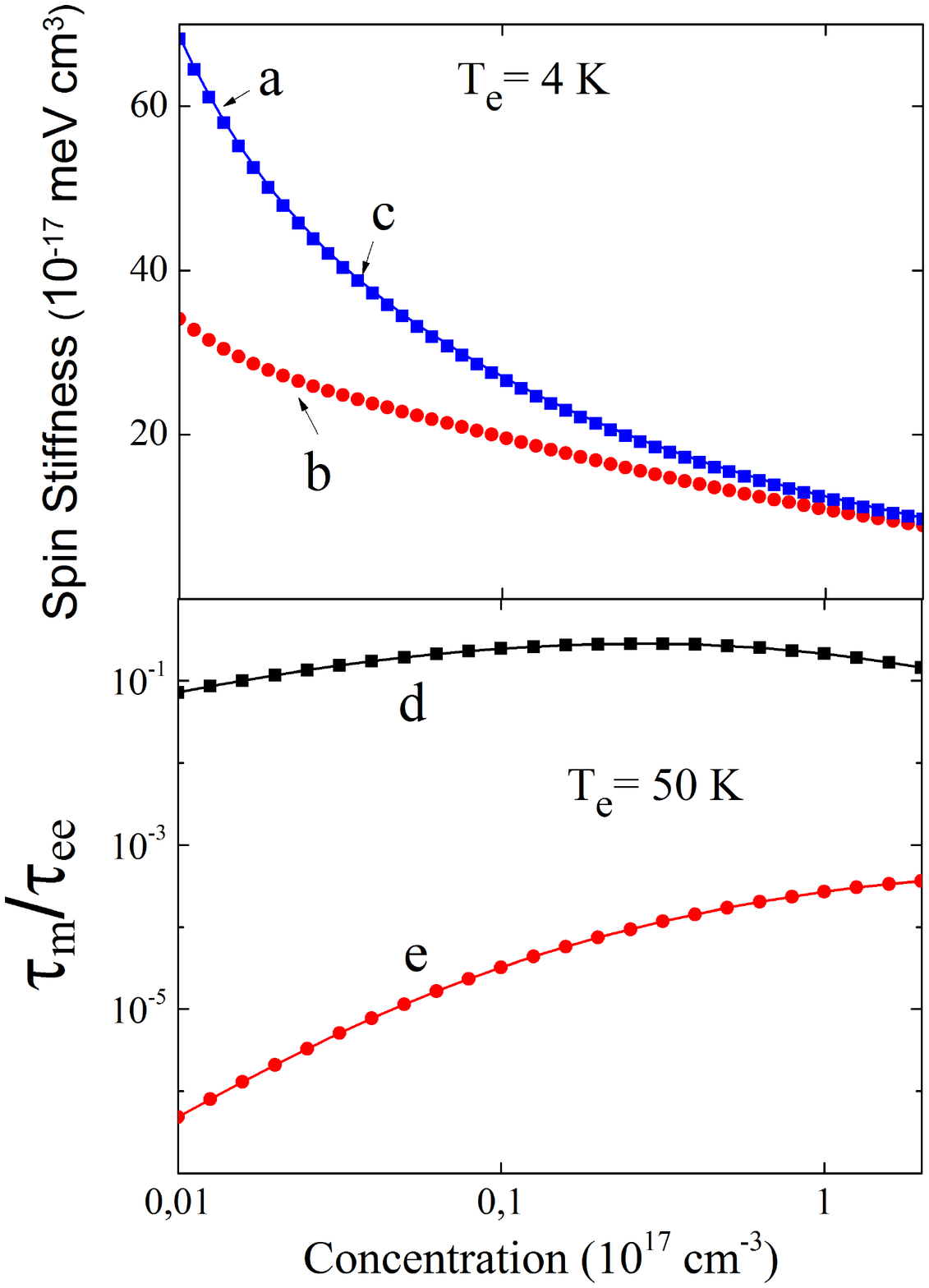}
\caption{Effect of electron-electron interactions on spin transport in $p^+$ GaAs. The top panel shows the dependence as a function of $n_i$ of the spin stiffness $S_{ii}$ [see Eq. (\ref{kimatrix})] without exchange interactions (Curve a), while the full circles (Curve b) and the full squares (Curve c) show the same quantity for  unscreened exchange interactions and screened interactions, respectively [see Eq. (\ref{stiffnesstot})]. The bottom panel shows the effect of screening on the efficiency of Coulomb spin drag. Curve d shows the quantity $\tau_m/\tau_{ee}$, calculated in Appendix C, as a function of the electron concentration without any screening by holes, and corresponds to results of  Ref. \cite{vignale2002}. The screening by holes of concentration $N_A^-= 10^{18}$ cm$^{3}$  is included in Curve e. The strong screening-induced decrease of $\tau_m/\tau_{ee}$ shows that, for the present material, spin drag is negligible. }
\label{spindrag}
\end{figure}

Such reasoning is applied to the calculation of the spin stiffness $S_{ii}$ in the presence of  electron-electron exchange interactions (bandgap renormalization). This calculation is detailed in Appendix C to first order and the result is given by Eq. (\ref{stiffnessexch}) in the low temperature limit.  Curve a of the top panel of Fig. \ref{spindrag} shows the spin stiffness dependence on the electron concentration $n_i$ at 4K.  Curve b and Curve c show the same quantity for unscreened and screened [$N_A+ \delta p = 10^{18}$ cm$^{-3}$] exchange interactions, respectively. In the absence of screening, exchange interactions induce a significant decrease of the spin stiffness, and therefore of the diffusion constant, as seen from Eq. (\ref{D}). This decrease is however completely cancelled if screening is included, as seen from the perfect correspondence between Curve a and Curve c. This screening is found  to be extremely efficient, since a small value of $N_A+ \delta p = 10^{15}$ cm$^{-3}$ is sufficient to produce a complete screening \cite{note7}. Although this result has been obtained in the low temperature limit, the conclusion remains true at higher temperature since the order of magnitude of renormalization corrections tends to decrease \cite{note1}. It is thus concluded that, in the presence of screening, electron-electron interactions are completely negligible. 

In the case of Coulomb spin drag, the effect of hole screening on the quantity $\tau_m/\tau_{ee}$ appearing in Eq. (\ref{eqE1}) is calculated in Appendix C using the theory developped in Ref.  \cite{vignale2002, vignale2005}. The result is shown in the bottom panel of Fig. \ref{spindrag}. Without screening, in agreement with Ref. \cite{vignale2002}, the maximum value of $\tau_m/\tau_{ee}$ is small but not completely negligible. Its value is about 0.3 and is reached near $n= 10^{16}$ cm$^{-3}$ which corresponds to the limit of degeneracy. For $N_A^-= 10^{18}$ cm$^{-3}$, it is found that the efficiency of spin drag is decreased by three orders of magnitude. As a result, the effect of Coulomb spin drag is completely negligible.\

\section{Interpretation}
\label{results}

\subsection{Calculation of the polarization profiles}

In order to determine the relative importance of the various processes considered in Sec IV, we have solved numerically the system of Eq. (\ref{electrons}), Eq. (\ref{spins}), Eq. (\ref{holes}) and Eq. (\ref{Poisson}), using an approximate method described in Appendix B and taking for the front and back surface recombination velocities the very weak values in our low temperature conditions $ S= S'= 5 \times 10^{4} $ cm/s \cite{cadiz2014}. The polarization profiles were then calculated using Eq. (\ref{imagesum}) and Eq. (\ref{imagediff}). The parameters  used for the resolution were all determined independently so that no fitting procedure was used. Their values are given in the present section. \

\begin{figure}[tbp]
\includegraphics[width=8 cm, trim=1 1 1 400, clip=true] {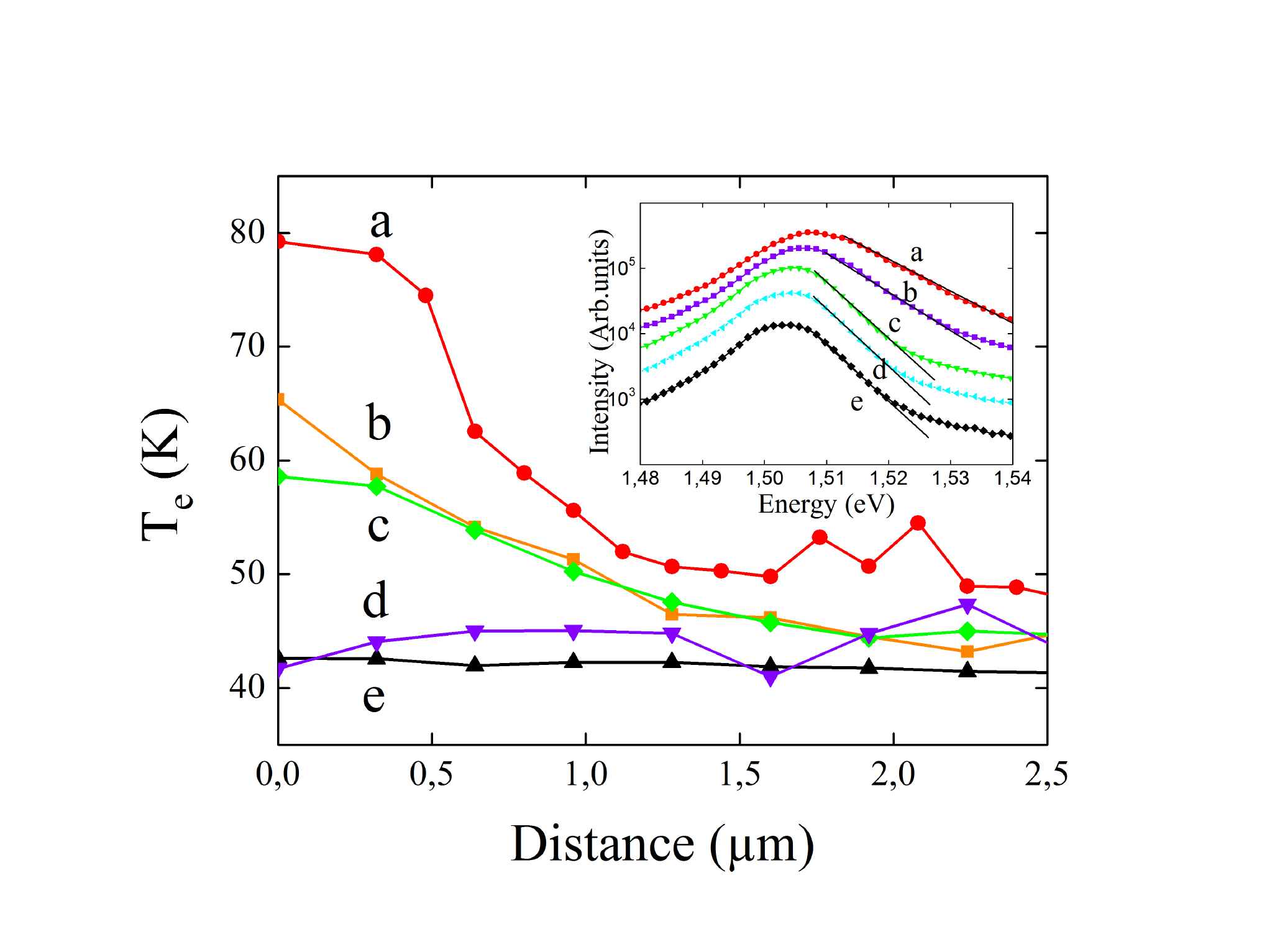}
\caption{The inset shows, for a large excitation power of 2.55 mW for $T=15K$, the spatially-resolved luminescence spectra at the place of excitation (a) and at a distance of 0.64 $\mu$m (b),  2.7 $\mu$m (c), 4 $\mu$m (d), and 9.6 $\mu$m (e). The larger electronic temperature $T_e$ at the place of excitation is evidenced from the high-energy side of the spectra. The main figure shows $T_e$ as a function of distance for different excitation powers : $2.55$ $mW$(a),  $1.89$ $mW$(b),  $1.03$ $mW$(c),  $0.45$ $mW$(d) and 1.5 $\mu$W (e).}
\label{T1}
\end{figure}

The increase of  the local temperature $T_e$ of the photoelectron gas caused by the increase of excitation power was first characterized. Shown in the inset of Fig. \ref{T1} are local luminescence spectra at high excitation power, as a function of distance to the excitation spot. The spectra exhibit a change in the shape of the high temperature tail, thus revealing a local heating of the electron gas near the place of excitation. It is assumed that the heating of holes is negligible \cite{leo1991}. Fig. \ref{T1} itself shows the spatial dependence of $T_e$ at a lattice temperature of 15K, for several excitation powers. At low power, the temperature is constant and equal to 40K. Conversely, at the maximum power, $T_e= 80$ K at the place of excitation and decreases to  50 K over a characteristic distance slightly larger than the radius of the laser excitation spot \cite{note4}. 

Hall effect measurements on an identical  contacted sample \cite{cadiz2014b} have shown that, at $T=15$ K, as in agreement with independent studies \cite{kim1997},  the concentration of ionized acceptors is $N_A^- \approx 10^{18}$ cm$^{-3}$, close to its value at 300 K. The hole mobility is $200$ cm$^2$/V.s. On the same contacted sample, the electron mobility was measured as a function of $T_e$ by monitoring the change of the luminescence profile induced by application of an electric field \cite{luber2006}. One finds $\mu_e=8800$ cm$^2$/V.s at  $T_e=50$ K and  $\mu_e=5800$ cm$^2$/V.s at  $T_e=75$ K. Using  Einsteins's relation at low power, this gives $D_0 \approx 37$ cm$^2$/s. The parameter $p$ defined in Eq. (\ref{s}) was estimated using a combined measurement of Hall and drift photoelectron mobility \cite{cadiz2014b}. One finds $p = 0 \pm 0.5$, which is close to the expected value $p = 1/2$ in the case where the mobility is determined by screened collisions with charged impurities or with majority holes \cite{smith1978,chatto1981}. This result implies that, for the present sample, the spin dependence of the mobility is weak. \

Finally, time-resolved polarized luminescence measurements as a function of $T_e$ were performed in the GaAs sample under consideration \cite{cadiz2014}.  At $T_e=50$K, one finds  $\tau_{eff} \approx$ 335 ps and $T_1 \approx$  1125 ps. While at $r=0$, $T_1$ can be smaller than the latter value, its value would still be much larger than the diffusion time [ Eq. (\ref{taueff0})], so that its possible decrease will very weakly affect the electronic polarization. As a result,  $T_1 \approx$  1125 ps was taken in all cases.\

\subsection{Discussion}
The calculated polarization profiles are shown in panel b of  Fig. \ref{fig05} for the same excitation powers as panel a. These profiles correspond quite well with the experimental results of panel a, apart from a slight difference in the position of the polarization maximum. As shown in Panel c  of Fig. \ref{fig05}, the polarization dip is restricted mainly to a zone labelled $\mathscr{D}$, defined by $z< 1\mu$m and $r< 0.3 \mu m$. Conversely, for $r \approx 1.5 \mu m$ and  $z<1\mu m$, the polarization maximum is as large as $42 \%$. At the highest excitation power, one calculates that the averages of the concentrations over $\mathscr{D}$  are $<n_+(\mathscr{D})>  \approx 9.0 \times 10^{16} \; \mbox{cm}^{-3}$ and $<n_-(\mathscr{D})> \approx 5.4 \times 10^{16}\; \mbox{cm}^{-3} $. These values are higher than the spin-resolved effective density of states at 80K which is $ N_{c}^s \approx 3 \times 10^{16} \; \mbox{cm}^{-3}$. With these concentrations, we calculate that $D_+=1.96 D_0$, $D_-=1.59 D_0$ so that $D_+/D_-= 1.25$. This value is slightly different from the experimental value of 1.49, which is not surprising because of the approximations used for the latter value.  \  
  
The importance of ambipolar diffusion is seen from Panel d of Fig. \ref{fig05}, which shows the profiles calculated in the unipolar case, by considering only Eq. (\ref{electrons}) and Eq. (\ref{spins}) and by taking $E=0$. It is striking to see that, in this case,  one observes only a small polarization dip in the profiles at the place of excitation. With the present values of $N_A$ and $n$, ambipolar diffusion results in reduced ambipolar diffusion constants $D^a_{cc}$ and $D^a_{cs}$, by the same amount $\beta_h$ defined by Eq. (\ref{kappa}). The latter quantity can be quite small since at low temperature $\mu _e/\mu _h \approx 85 $. This results in an increase of the concentrations and therefore of the amount of degeneracy at $r=0$.  In the unipolar case, we find at the highest excitation power $<n_+(\mathscr{D})> \approx 5.2 \times 10^{16} \; \mbox{cm}^{-3}$ and $<n_-(\mathscr{D})> \approx 2.2 \times 10^{16}\; \mbox{cm}^{-3} $. The total concentration is smaller than its above ambipolar value by a factor of $\approx 2$, which would approximately correspond to the experimental case for a reduced power of between 1.03 mW and 1.89mW. As seen in Panel a of Fig. \ref{fig05}, the unipolar calculated profile at 2.55 mW is indeed intermediate between the experimental profiled at the latter two powers.\

 Note finally that the concentration in the unipolar case is  still larger than $N_{c}^s$ so that some amount of degeneracy is still present. Indeed we calculate $D_+/D_-= 1.26$ i. e. quite similar to the value at high power. This is because, in Eq. (\ref{eta}), the decrease of $\delta$ caused by the smaller concentration is compensated by the polarization increase so that the actual value of $D_+/D_-$ only weakly depends on concentration. On the other hand, the actual value of the  polarization in $\mathscr{D}$ is the result of a self-consistent equilibrium and can be relatively sensitive to the concentration.\cite{note8}\

Panel e shows the profile calculated under the same conditions as Panel b, except that the Soret charge and spin currents are neglected. Here  $T_e$ is taken as spatially homogeneous and equal to its measured value at $r=0$. Apart from the highest power where the profile is slightly shifted upwards, the profiles are nearly the same as in panel b implying that the Soret current plays a negligible role in these experiments \cite{note5}. The relative temperature gradient $\vec{\nabla}_r T_e/T_e$ strongly varies with distance. As found from Curve a of Fig. \ref{T1}, its value at high power is very small near $r=0$, reaches 1.3 $\mu$m$^{-1}$ in a very short interval near $0.6 \mu$m, and then decreases to $0.3 \mu $m$^{-1}$. In comparison, the relative charge gradient $\vec{\nabla}_r n_e/n_e$, found using Fig. \ref{figexp},  is almost independent of distance and is of the order of $1 \mu $m$^{-1}$ and is, within experimental uncertainties, larger than the temperature relative gradient at most distances. Using Eq. (\ref{compsoret}), it is thus concluded that the temperature gradient is not sufficient to obtain significant Soret currents. 

Panel f of Fig. \ref{fig05} shows the polarization profile calculated by considering the unipolar limit without temperature gradients, $T_e$ being fixed to its measured value at $r = 0$. This situation is reminiscent of spatially homogeneous configuration of spin grating experiments. In this case the polarization dip near $r = 0$ has almost disappeared. Observation of the Pauli blockade driven spin filter effect thus requires spatially inhomogenous electron and hole concentrations. This means that the usual spin grating technique, in which the electron and hole concentrations are uniform in space, may not be well adapted to the observation of Pauli-blockade effects in spin transport. On the other hand, for spin gratings, $T_e$ is also uniform in space, meaning that the charge and spin Soret effects are absent, a situation which should slightly increase the magnitude of Pauli-blockade phenomena. Given that heating of the photoelectron gas is unavoidable during high intensity photoexcitation, the ideal conditions for measuring the largest possible Pauli blockade effects are highly inhomogenous photoelectron and hole concentrations and spatially uniform temperatures.\

\section{Conclusion}
\label{conclusion}
Here, we present a theoretical and experimental investigation of the effect of degeneracy on spin transport of a photoelectron gas. We have used $p^+$ GaAs for which at 15K tightly-focussed circularly-polarized light excitation generates strongly spin-polarized  photoelectrons ($45 \; \%$) and where charge and polarization profiles are monitored as a function of distance. We now recall the main results : \

a) In conditions where the photoelectron gas is degenerate, i. e. for a sufficiently low temperature and large excitation power (above 1mW), we demonstrate a novel spin-charge coupling mechanism implying a spin-dependence of the diffusive transport, with relative differences in the spin-resolved diffusion constants as large as $50\%$ between the two types of spins. This effect is linear in the electronic polarization, increases with the electron concentration and decreases with increasing kinetic energy in the conduction band. The spin-averaged charge diffusion constant is also shown to be spin-dependent due to a second order effect.  The dominant effect which explains these results is the charge and spin dependence of the spin stiffness under degeneracy.\ 

b) Ambipolar diffusion plays a key role for the observation of spin-dependent diffusion,  since it increases the confinement of photoelectrons at the place of excitation, and therefore the amount of degeneracy, due to the electrostatic electron-hole coupling. This diffusion induces a strongly nonlinear coupling between electron diffusion, spin diffusion, and hole diffusion, which is treated  here using an approximate resolution of the diffusion equations. Such ambipolar-induced increase of the confinement could also be obtained by increasing the excitation power, but this will inevitably increase the electron temperature and decrease the degeneracy.\

c) The mobility is predicted to depend on charge and on spin. However, for $p^+$ GaAs, this effect is weak. This conclusion is at variance with the hypothesis of the previous work \cite{cadiz_prl2013} and is based on recent measurements of the dependence of the scattering time on kinetic energy \cite{cadiz2014b}, as defined by the value of $p$ in Eq. (\ref{s}). The value of $p$ is found to be strongly reduced from  $3/2$ due to scattering with charged impurities, as a consequence of the screening by holes. It is anticipated that the spin-dependence of the mobility should be observable at a lower p-type doping.\ 

d) Since the electronic temperature is strongly inhomogeneous,  thermoelectric currents may appear due to the Soret effect, which are predicted to depend on spin in degenerate conditions. However, in the present situation, this dependence does not strongly affect the polarization profile because of the relative values of charge and temperature gradients. Distinct experimental configurations should be used for separate investigation of this effect. 

e) Other spin-spin or spin-charge coupling mechanisms such as spin drag or bandgap renormalization are negligible in the present case because of efficient screening of the electron-electron interactions by the holes of our $p^+$ material.  
 
In summary, the extensive theoretical analysis of the present work and the careful sample characterization allow us to conclude that we have achieved experimentally a relatively simple situation, where the polarization profiles mostly depend on spin-dependent ambipolar diffusion under degeneracy. It is predicted that other effects could play a role under degeneracy such as spin-dependent mobility or Soret currents. These effects remain unobserved and could be explored by adjusting the acceptor density and the laser energy and power. Note that the present technique relying on a tightly-focussed laser excitation, seems better adapted than the elegant spin-grating technique for investigating the effect of degeneracy on spin transport. The main reason is that effects of degeneracy on pure spin currents created in the latter technique are not amplified by ambipolar diffusion. They decrease under increase of excitation power because of the unavoidable heating of the electron gas which reduces the degree of degeneracy.\

It is finally  pointed out that these large spin-dependent effects have been observed in a regime near the onset of degeneracy, where the photoelectron concentrations are not very large with respect to the effective density of states in the conduction band. This implies that much stronger effects are expected for larger powers. While this is not possible in the present case because of heating effects, we anticipate that the use of appropriate low dimensional structures of reduced effective density of states will increase the magnitude of the effects and may possibly open the way to the realization of spin components of increased diffusion length and mobility at a  temperature closer to $300$ K.  

\acknowledgements{We are grateful to J.-P. Korb and D. Grebenkov for valuable help in the  characterization of the sample dynamic properties. We thank E. L. Ivchenko and S. A. Tarasenko  for extremely fruitful discussions. One of us (F. C.) is grateful to CONICYT Grant Becas Chile for supporting his work.}

\appendix

\section{Boltzmann equation formalism for the charge, spin and thermoelectric currents}
\label{appendix B}

The current $\vec J_i$ of photoelectrons of  spin $i$ is given by 
\begin{equation}
\vec J_i = -\frac{q}{m^*}\int \vec p f_i d^3 p
\label{A1}
\end{equation}
\noindent
where the function $f_i$,  which describes the distribution of electrons of spin $i$ as a function of space and of momentum $\vec p$, is obtained from a resolution of the Boltzmann equation   
\begin{equation}
\frac{\partial f_{i}}{\partial t} + \vec p \vec \nabla_r f_{i}- \frac{q\vec E}{m^*} \nabla_k f_{i}=[\frac{\partial f_{i}}{\partial t}]_{i coll}+[\frac{\partial f_{i}}{\partial t}]_{e-e coll}
\label{Boltzmann}
\end{equation}

\noindent
where the second term of the left hand accounts for the effect of diffusion in a Fermi energy gradient. The third term describes the effect of electric field and the two terms on the right hand are collision integrals accounting for electron-impurity collisions and electron-electron collisions. 

The models of Refs. \cite{vignale2002, Flensberg2001, glazov2004} propose estimates of the collision integrals, but do not take into account spatial inhomogeneities of $f$. These inhomogeneities are considered in an independent approach, which  however neglects the  spin polarization, so that electron-electron-collisions have no effect \cite{chakravarti1975}. Here, neglecting band nonparabolicity, we propose the following Ansatz to first order which reduces to the result of  Ref. \cite{vignale2002} for a homogeneous electron gas and to that of Ref. \cite{chakravarti1975} for spin-unpolarized electrons. 	
\begin{align} \label{A4} f_i = f_{0i}&-\frac{\alpha_{ii}\tau_{m}(\varepsilon)}{m^*}\left[-q\vec E \cdot \vec \nabla_{\varepsilon} f_{0i} + \vec p \cdot \vec \nabla_r f_{0i} \right] \nonumber \\ &-\frac{\alpha_{i, -i}\tau_{m}(\varepsilon)}{m^*}\left[-q\vec E \cdot \vec \nabla_{\varepsilon} f_{0, -i} + \vec p \cdot \vec \nabla_r f_{0, -i} \right]  \end{align}	 
\noindent
where for non coupled spins ($\alpha_{i, -i}=0$) one recognizes the usual drift term in the electric field $\vec{E}$ and the diffusion term proportionnal to the spatial gradient $\vec \nabla_r f_{0i}$ \cite{chakravarti1975}. In order to take account of the spin-spin interactions for the evolution of $f_{i}$, it is natural to add a coupling term with the evolution of $f_{-i}$, using the same coupling factor  $\alpha_{i, -i}$ as the one given by Eq. (\ref{eqE1}), which describes the modification of conductivity $\sigma_{i, -i}$ caused by e-e collisions. In the same way, the evolution of $f_{i}$ is also modified by losses to the $-i$ spin system, which are taken into account by the muliplicative factor $\alpha_{ii}$. It is considered here that $\tau_{m}(\varepsilon)$ does not depend on spin, since the spin dependence of $\tau_{mi}$ used in Sec. II originates from the sole spin dependence of the Fermi distribution. Eq. (\ref{A1}) allows us to calculate the currents using Eq. (\ref{A4}) and 

\begin{align}
	 \vec \nabla_r & f_{0i}=-\frac{\partial f_{0i}}{\partial \varepsilon} \cdot \nonumber \\ & \left[ \sum_j \frac{\partial E_{F_i}}{\partial n_j}\vec \nabla_r n_j+(\frac{\partial E_{F_i}}{\partial k_BT_e}+\frac{E-E_{F_i}}{k_B T_e})\vec \nabla_r (k_B T_e) \right]
	 \label{A6}
\end{align}			
\noindent
Since the contribution of  the equilibrium term $f_{0}$  is zero, the current is written as the sum of a drift current, of a diffusion current and of a thermoelectric current, respectively proportional to $\vec{E}$, $\vec \nabla_r n$, and $\vec \nabla_r (k_B T)$. This gives the expressions of the drift and diffusion currents given in  Sec. IIIA1. Transforming the integration over momentum to an integration over kinetic energy, the expression of the average time $\tau_{mi}$ given by Eq. (\ref{taumi}) is readily obtained. The thermoelectric charge and spin currents originate from the second term of Eq. (\ref{A6}). The thermal-induced change of $E_{F_i}$ at constant concentration, $\partial E_{F_i}/\partial k_BT_e$ is calculated by expressing that the derivative of $n_i$ with respect to temperature, as found from Eq. (\ref{Nc}), is zero. Using $\partial  \mathscr{F}_{k}^*(\eta)/\partial \eta=\mathscr{F}_{k-1}^*(\eta)$, one finds the expression given in Eq. (\ref{matrixK}) for $\vec K_{ij}$. \\

\section{Solution of the equations for ambipolar spin diffusion}
\label{appendix B}

In contrast with the usual treatments of ambipolar diffusion \cite{zhao2009,smith1978}, the system of Eq. (\ref{electrons}), Eq. (\ref{spins}), Eq. (\ref{holes}) and Eq. (\ref{Poisson}) must be solved numerically since the conductivities and diffusion constants depend on space. However, an exact numerical solution of these equations is difficult, since small errors in $n$ and $\delta p$ results in large errors in $\vec E$. This renders the equations highly nonlinear and a convergent solution is difficult to obtain using finit element methods without approximations. To address this, the hole continuity equation is replaced by a combination of  combination of Eqs. (\ref{electrons})  (multiplied by $\sigma_h$) and (\ref{holes}) (multiplied by $\sigma_c$) in the usual way \cite{smith1978}. Defining the reduced hole conductivity $\beta_h=\sigma_h /(\sigma_h + \sigma_c)$, of the form 
 \begin{equation}   
  \beta_h=\frac{N_A^- + n}{(N_A^- + n) + (\mu_0/\mu_h)[n_+ \zeta (n_+)+n_- \zeta (n_-)] }  
 \label{kappa} 
 \end{equation}
the following equation to describe the hole distribution is obtained

\begin{align}    (g_++g_-)-& \delta p/\tau + \frac{\vec E}{q} \vec \nabla \sigma_{c}^a\nonumber \\ & +\vec \nabla [D_{cc}^{a}\vec \nabla \delta p  + D_{cs}^{a}\vec \nabla s + \frac{1}{q} \beta_h \vec \nabla \vec J_{c}^{T}]=  0 \label{charge_amb} \end{align}
 where 
 \begin{equation}    D_{cc}^{a}= \beta_h D_{cc} + (1-\beta_h) D_h \label{Da}\end{equation}
 \begin{equation}    D_{cs}^{a}= \beta_h D_{cs} \end{equation}
 \noindent
 and $ \vec \nabla \sigma_{c}^a =\beta_h \vec \nabla \sigma_{c}-(1-\beta_h) \vec \nabla \sigma_{h}$.\   
Equation (\ref{charge_amb}) is approximate since, as justified in Ref \cite{paget2012}, it assumes charge neutrality [$n=\delta p$]. Further, it neglects for simplicity the spatial dependences of electron and hole conductivities. However, this approximation appears to yield reasonable results. For example, at the highest excitation power where the equations are most strongly coupled, the sum of all the terms of the left hand of Eq. (\ref{electrons}) is two orders of magnitude smaller than the maximum value of  $\vec \nabla \cdot [D_{cc} \vec \nabla n]$ so that these terms efficiently compensate each other.\

\section{Effect of screening by holes on electron-electron interactions}
\label{appendix C}

a) We first estimate the contribution of electron-electron interactions  to the electron spin stiffness in the presence of a degenerate hole gas. The electron mutual interactions in the electron gas lead to a self energy correction to the bare electron energy, usually split into exchange and correlation terms and given by $\Sigma_{i,k}=\Sigma_{i,k}^x+\Sigma_{i,k}^{cor}$.   The effective Fermi energy is then given by $E_{Fi}^*=E_{Fi} +\langle \Sigma_{i,k}^x \rangle$ where  $\langle \rangle$ denotes the average over all electrons of spin $i$ and the contribution of many-body effects to the spin stiffness is thus given by 

\begin{align}
\label{stiffnesstot}
 S_{i,j}^{xc}= \frac{\partial \langle \Sigma_{i,k}^x \rangle}{\partial n_j}  \delta_ {i,j}+ \frac{\partial \langle \Sigma_{i,k}^{cor} \rangle}{\partial n_j}  
\end{align}	 
since the exchange correction is computed within a population of electrons of the same spin $i$. In degenerate conditions, we will neglect the  contribution of the correlation energy, which  is small with respect to the exchange one \cite{note1} and the spin stiffness matrix is diagonal. At low temperature, one has $\Sigma_{i,k}^x =-(1/V) \sum_{q} v(q) f_{i, k+q}$ where $V$ is the sample volume and $v(q)$ is the screened potential given by  Eq. (\ref{potential}). The exchange energy for electrons of momentum $k$ and spin $i$ can be analytically computed as  
\begin{align}
\label{exchangedef}
  \Sigma_{i,k}^x=-\frac{e^2 k_{Fi}}{\epsilon \pi} \mathscr{B}(\frac{k}{k_{Fi}},\frac{k_{DH}}{k_{Fi}})
\end{align}
where the negative function $\mathscr{B}$ is given by  
\begin{align}
\label{B}
   -\mathscr{B}(y_1, y_2)=  1+& y_2  \arctan \frac{y_1-1}{y_2}  +y_2  \arctan \frac{y_1+1}{y_2} \nonumber \\ & + \frac{1 + y_2^2- y_1^2}{4y_1} \ln \frac{(1+y_1)^2+y_2^2}{(1-y_1)^2+y_2^2}
\end{align}
and reduces to the usual expression \cite{mahan1981, sun2010} in the absence of holes.  After integration over all electrons of spin $i$, the average exchange energy is written
\begin{align}
\label{exchangeaverage}
  \langle \Sigma_{i,k}^x  \rangle = \frac{3 e^2 k_{Fi}}{2 \epsilon \pi} \int^1_0 y_1^2 \mathscr{B}(y_1,\frac{k_{DH}}{k_{Fi}}) dy_1
\end{align}
and the additional spin stiffness is given by 
\begin{align}
\label{stiffnessexch}
 S_{i,i}^{x} = \Big[\frac{3}{4 \pi} \Big] \frac{e^2}{ \epsilon } \int^1_0 y_1^2 \mathscr{R}(y_1,\frac{k_{DH}}{k_{Fi}}) n_i^{-2/3} dy_1
\end{align}
where 
\begin{align}
\label{R}
 -\mathscr{R}(y_1, y_2)=  1+\frac{1 + y_2^2- y_1^2}{4y_1} \ln \frac{(1+y_1)^2+y_2^2}{(1-y_1)^2+y_2^2}
\end{align}

b) We now estimate the effect of screening on spin drag. The relative efficiency of spin drag under screening is measured from the ratio $\tau_m/\tau_{ee}=\rho_{+-} \sigma_c$ where the spin transresistivity $\rho_{+-}$ is expressed as an integration over frequency followed by an integration over momentum. \cite{vignale2002,vignale2005}

\begin{align}
\label{transres}
 \rho_{+-} = &\frac{\hbar^2}{q^2n_+n_-}\frac{1}{3\pi^3} \int^\infty_0 k^4 v(k)^2 dk \nonumber \\ & \int^\infty_0 d\omega \frac{  \chi_{0+}^"(k, \omega) \chi_{0-}^"(k, -\omega)}{\left|\epsilon(k, \omega)\right|^2 \sinh^2( \hbar \omega/2k_BT)} 
\end{align}	 
	 
\noindent
where  the dynamic dielectric constant is given by 
\begin{equation}   
  \epsilon(k, \omega)=1 - v(k)[\chi_{0+}(k, \omega)+ \chi_{0-}(k, \omega)] 
 \label{epskw} 
 \end{equation}

\noindent 
Here, $\chi_{0i}(k, \omega)$ is the noninteracting spin-resolved density-density response function of spins $i$ and $\chi_{0i}^"(k, \omega)$ is its imaginary part.  The expressions of the latter quantities can be found in Ref. \cite{vignale2005}. Since $\chi_{0i}^"(k, \omega)$ does not directly depend on the potential, it is natural to include the effect of screening by holes by  using for the potential $v(k)$, the expression given by Eq. (\ref{potential}).  

\bibliographystyle{apsrev}


\begin{thebibliography}{51}
\bibitem[{not({\natexlab{a}})}]{note3}
\bibinfo{note}{This value is slightly lower than that of the two preceding
  subsections, probably because of a slightly smaller electronic polarization
  caused by a slightly increased electron temperature}.
\bibitem[{not({\natexlab{b}})}]{note7}
\bibinfo{note}{This result should only be taken qualitatively, since in this
   case, the electron concentration is larger than the hole one so that
  screening by electrons should also be taken into account in a calculation to
  second order}.
\bibitem[{not({\natexlab{c}})}]{note1}
\bibinfo{note}{We have checked, using Refs \cite{sun2010, brown2013} that, for
  $T_e = 60 K$ and $n = 10^{16} cm^3$, the exchange term dominates the electron
  gas energy, and can be approximated by its value at $T_e=0K$.}
\bibitem[{not({\natexlab{e}})}]{note5}
\bibinfo{note}{Since the spin dependence of the Soret current is the same as
  that of the diffusive current [see Eq. (\ref{ratioK}) for $\theta=1$],
  inclusion of the Soret current essentially does not modify $s$. On the other
  hand, at the highest power, there occurs a slight decrease of the
  concentration by the Soret current, which increases the polarization.}
\end{thebibliography}

\end{document}